\documentclass[a4,11pt]{report}
\pdfoutput=1
\usepackage{graphicx}
\usepackage{epsfig}
\usepackage{listings}
\usepackage{subfig}
\usepackage{caption} 

\makeatletter

\def\cs{$\chi^2 \;$} 

\begin{document}

\title{Chameleon: A Reconstruction package for a KM3NeT detector}

\author{ D. Lenis and G. Stavropoulos }

\maketitle
\tableofcontents
\newpage
\section{Introduction}
In this note we describe the Chameleon software we developed for the
event reconstruction of KM3 detector.
This software package's  developement started  as a standalone
application before the endorcement from the KM3NeT consortium of the
SeaTray software framework, but it was adapted to it on the course.

Chapter 1 outlines the techniques we developed for the pattern
recognition and the track fitting.
 In Chapter 2, we demonstrate the performance of the Chameleon Reconstruction.

\chapter{Pattern Recognition and Fitting}

There are two main parts in the reconstruction package: The
patern recognition and the fitting.
The first part is designed for a km$^3$ multi PMT Optical
Modules detector (see e.g. \cite{elswriteup}). It consists of a
algorithms for the  selection and grouping of
hits in track canidates.
 
The second is a generic $\chi^2$ minimizer. This 
is generic enough to
allow for unbiased comparison between different detector designs. 

Before the description of the the pattern recognition and fitting
algorithms we give  a short description of the specific data sets used for
this report.

\section{Data Sets}
\label{sec:data}

The demonstration of the reconstruction is done on a sample of  6,000,000  
MC neutrinos produced through the de facto standard tools provided
with the seatray, namely {\tt nugen} for the production and {\tt g4sim} for the
simulation.  
\begin{figure}[!htpb]%
  \centering
  \includegraphics[width=70mm]{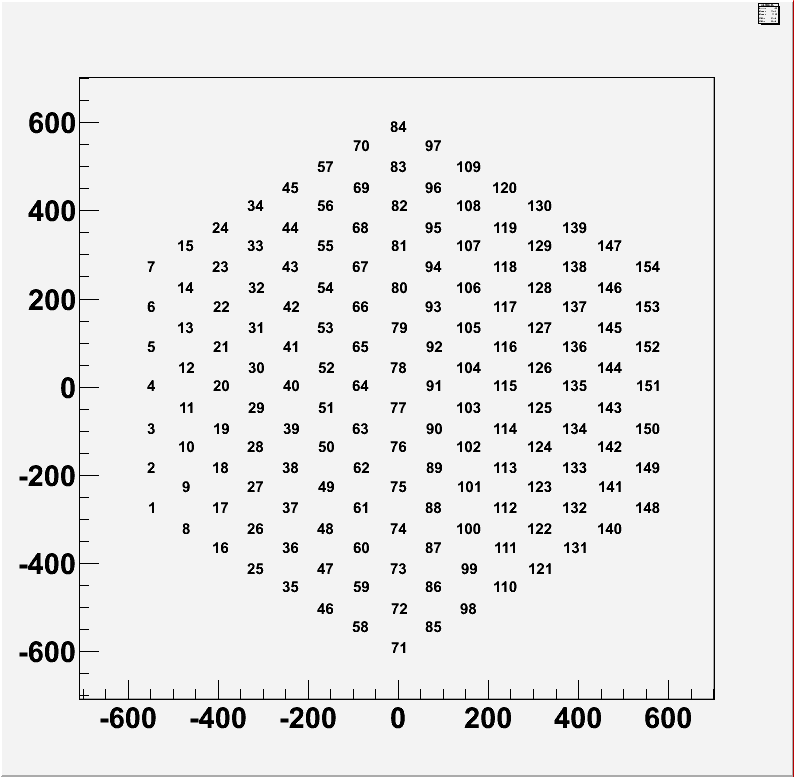}
\caption{ The detector geometry used in this study.}
\label{fig:detector}
\end{figure}
In this data set, no light scattering in water was simulated.
$^{40}K$ noise was simulated at a rate of $8 KHz$ per PMT, which is
more than the commonly accepted rate for these tubes of $5.2 KHz$, since 
more noise would be a stringent test of the filtering out of the
noise.
For a detailed description of the whole simulation, see \cite{claudio}.

The detector used is a small, dense detector of 154 strings
(fig. \ref{fig:detector}), with string distance 92 m. Each string has
an active length of 570 m., and 20 Multi PMT
Optical Modules at 30 m. distance between each other. Each
OM is equipped with 31 $3^{\prime\prime}$ PMTs.

\begin{figure}[!htpb]%
  \centering
  \subfloat[Generated $\nu$ energies]
  {\includegraphics[width=90mm]{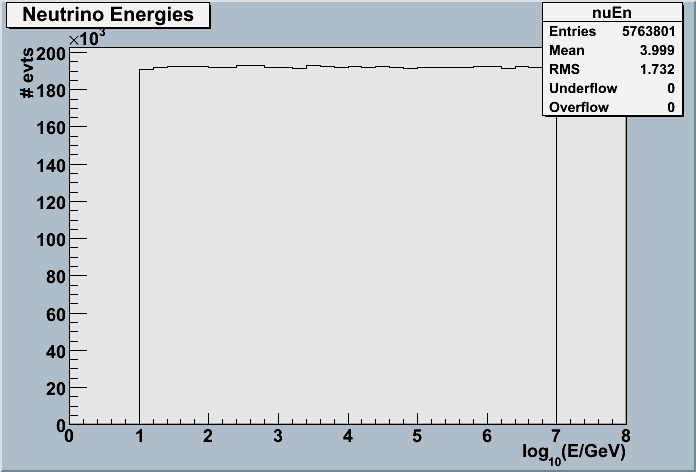}}\\
  \subfloat[$\theta$ and $\phi$ angle distribution]{
  \includegraphics[width=90mm]{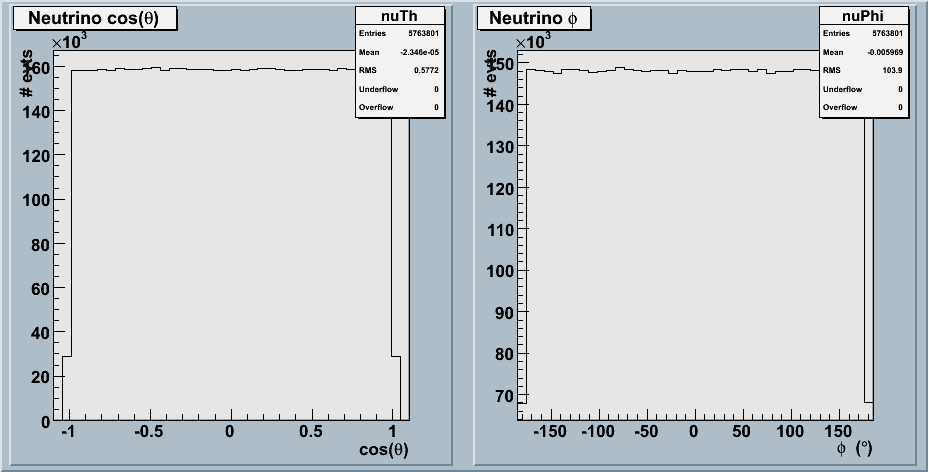}}
  \caption{Generated $\nu$ spectra.}
  \label{fig:nudistr}
\end{figure}

\begin{figure}[!htpb]%
  \centering
  \subfloat[Generated CC $\mu$ energies]
  {\includegraphics[width=90mm]{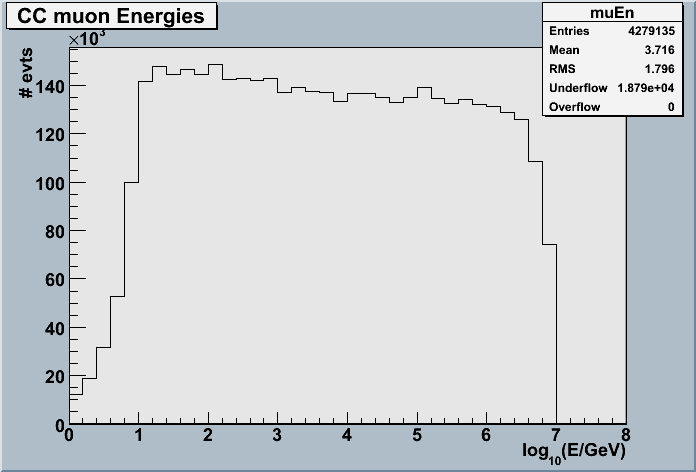}}\\
  \subfloat[$\theta$ and $\phi$ angle distribution]{
  \includegraphics[width=90mm]{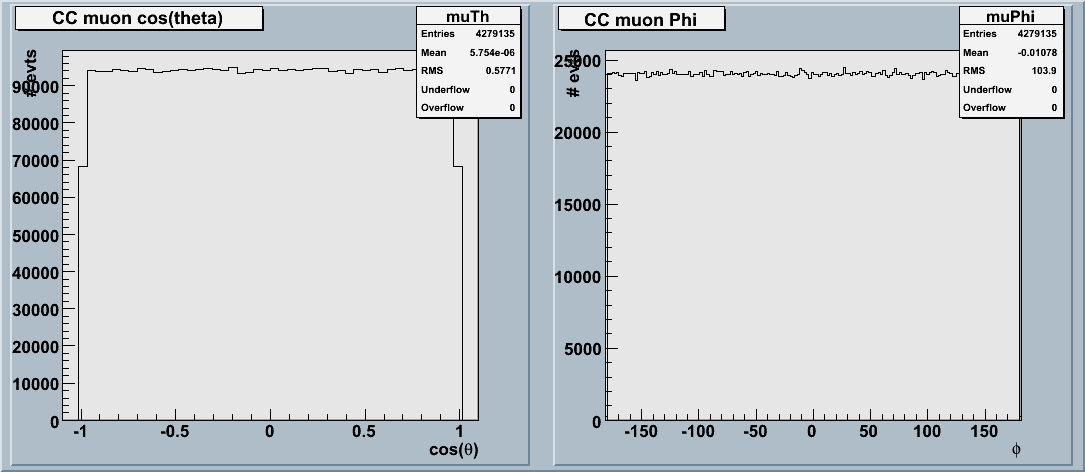}}
  \caption{Generated CC $\mu$ spectra.}
  \label{fig:mudistr}
\end{figure}

The produced $\nu$-spectrum, which corresponds to an $E^{-1}$ flux, is presented in fig. \ref{fig:nudistr}, 
while the respective quantities for charged current produced muons in 
fig. \ref{fig:mudistr}.

\begin{figure}[!htpb]%
  \centering
  \subfloat[Distribution of muons as a function of zenith ~$z$]
  {\label{fig:zenithdistri}
    \includegraphics[width=75mm]{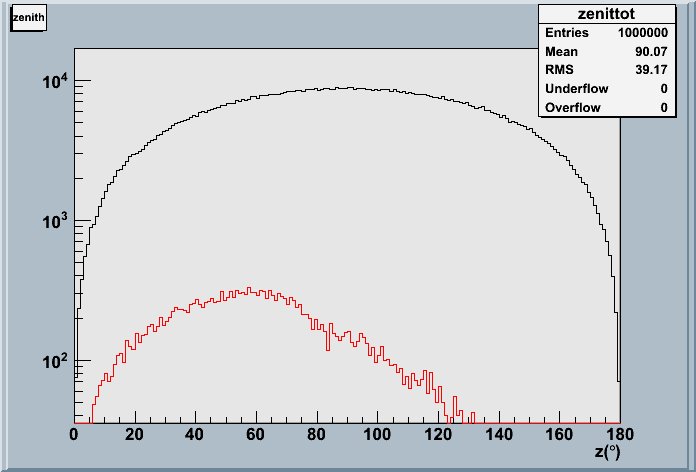}}\quad
  \subfloat[Distribution of muons as a function of  $\cos(z)$]
  {\label{fig:zenithcosdistri}
    \includegraphics[width=75mm]{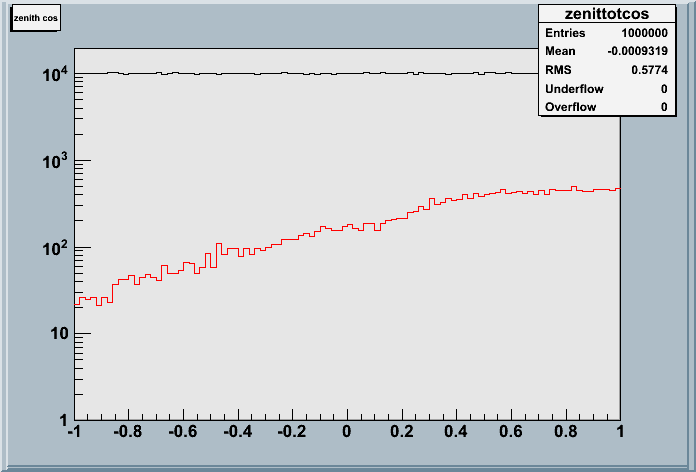}}\\
  \subfloat[Distribution of muons as a function of $\alpha$]
  {\label{fig:azidistri}
    \includegraphics[width=75mm]{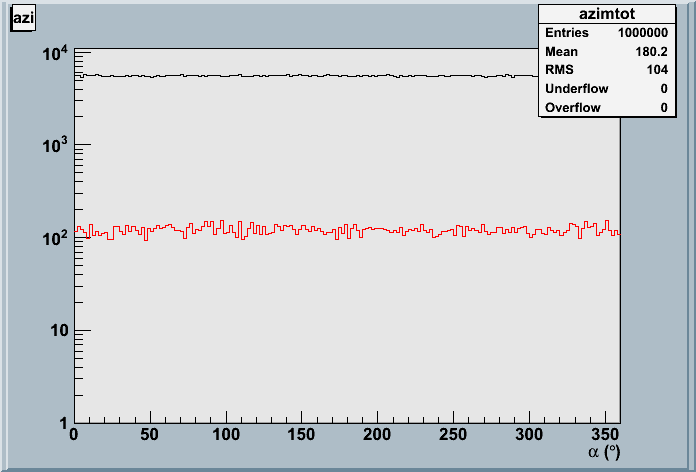}}
  \caption{Angle distributions for $\mu$'s.: Black all $\mu$'s, Red reconstructed $\mu$'s}
  \label{fig:muangledistri}
\end{figure}

\begin{figure}[!htpb]%
  \centering
  \subfloat[CC events]
  {\includegraphics[width=90mm]{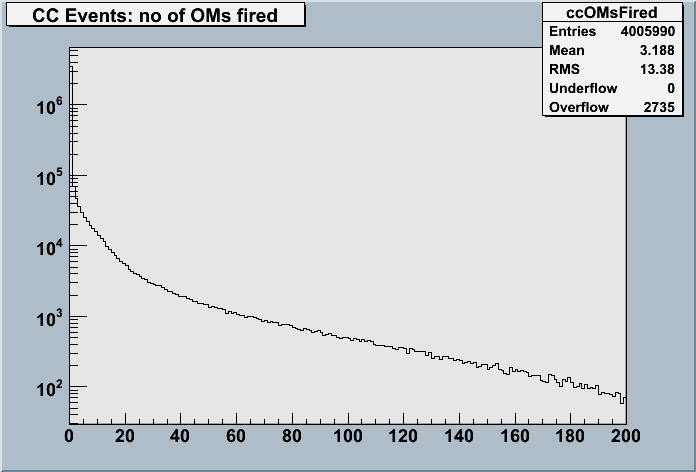}
   \label{fig:ccoms}}\\
  \subfloat[NC events]{
  \includegraphics[width=90mm]{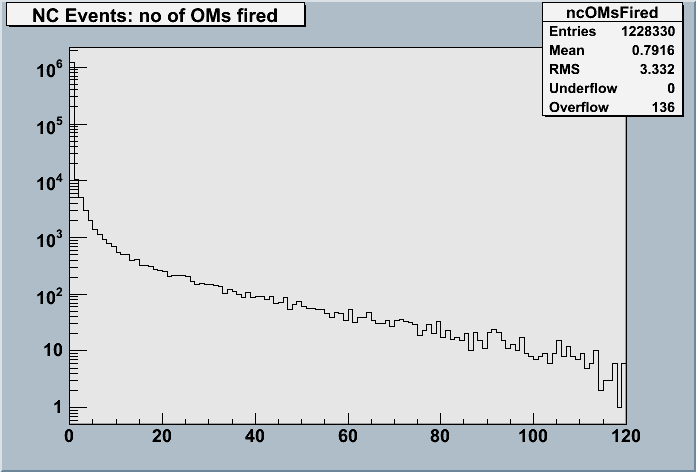}
  \label{fig:ncoms}}
  \caption{Numbers of OMs fired.}
  \label{fig:numofoms}
\end{figure}

A measure of the total number of produced photons is the number of OMs
that  fired, which is shown in fig. \ref{fig:ccoms}, for CC events and
in \ref{fig:ncoms} for NC events. CC events are defined by the
existence of a primary muon in the particle tree, while NC events by
the absence thereof.


The flat input neutrino spectrum is transformed into 
the relevant fluxes with the use of the ``NeutrinoFlux'' module
(implemeneted originaly for IceCube). Specifically we use  functions 
{\tt  ConventionalNeutrinoFlux("bartol\_numu")} and \\{\tt
  PromptNeutrinoFlux("naumov\_rqpm\_numu")}. The first simulates the
conventional muon flux (leptons from $\pi^{\pm}$'s and Kaons), while
the second simulates leptons
from charm decays. The atmospheric flux is
the sum of these two functions. 


\section{ Fitting}
\label{sec:fit}
An integral part of the pattern recognition are the methods developped
for the fitter, so a description of the fitting algorithm will
necessarily precede the description of the hit selection and final reconstruction. 

\begin{figure}[!htpb]
\centering
\includegraphics[width=90mm]{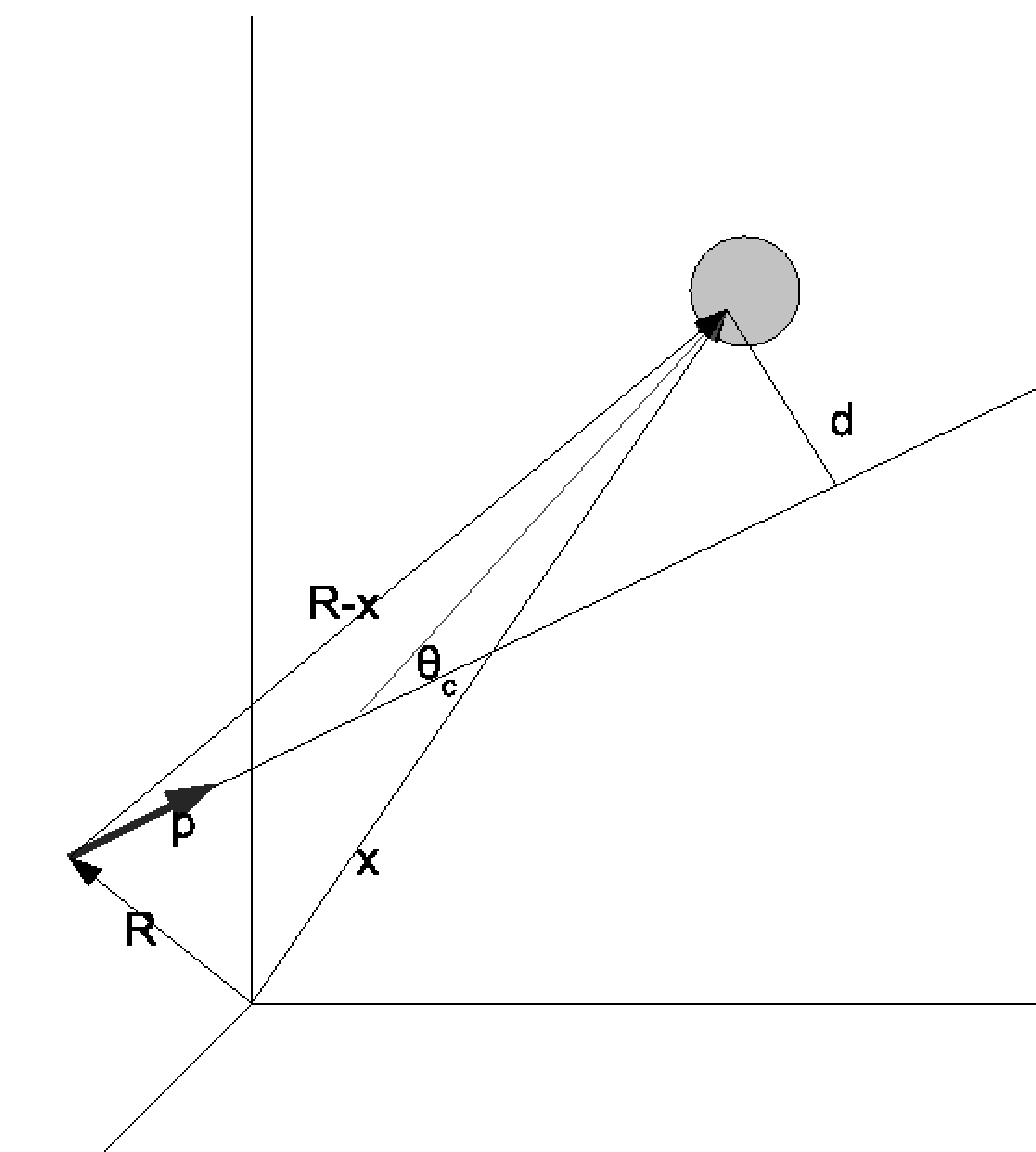}%
\caption{The geometrical model used for the minimization. The \v{C}erenkov index used is
for $\lambda = 460 nm$ ($43.5^{\rm o}$). }
\label{fig:geom}
\end{figure}

The fitter is an implementation of a \cs minimization with the help of
the {\tt C++ Minuit2} package (embedded within ROOT).
The minimization is done according to the geometrical model of fig. \ref{fig:geom}.

Specifically, if a muon with  (pseudo) 
vertex at $\bf{R}$ and momentum direction $\bf{p}$, emits a photon that hit 
a PMT at $\bf{x}$, the total time between the vertex and the hit is given by
\begin{equation}
 c\;t^{exp} = \sin \theta_c \sqrt{(\bf{x} - \bf{R})^2 -\left[ \bf{p}\cdot  (\bf{x}-\bf{R})\right]^2 }
+ \bf{p}\cdot(\bf{x}-\bf{R}),
\label{eq:geom}
\end{equation}
where $\theta_c$ the \v{C}erenkov angle. The \cs function that is
minimized is
\begin{equation}
  \label{eq:chi2}
  \chi^2 = \sum_{i}\frac{t_i^{exp} - t_i^{data}}{\sigma_i},
\end{equation}
where $t_i^{exp}$ is the expected arrival time of the $i$th photon of
eq. (\ref{eq:geom}), assuming that the measured pulse is the PMT response to
\v{C}erenkov light originating from a muon track,  $t_i^{data}$ is the
actual measured time, and $\sigma_i$ is the error ascossiated with the
$i$th hit. This error is set equal to 2 ns.

Within the project, two distinct strategies for the estimation of
track parameters
were implemented. They are based on an assesment of the hit  residuals
wrt the track, as it is produced by the fit.
\begin{itemize}
\item {\em Fit with Rejection:} This is a recursive  fit, which
  calculates the hit residuals and rejects 
  those hits that are $N\sigma$ 
  away from the nominal  value of $2~ns$. 
\item {\em Constant deweight:} All hits are deweighted and their
  $\sigma$ is set equal to a constant parameter.
\end{itemize}

The fitter is able to calculate the track parameters  in either
cartesian coordinates (track vertex ${\bf R}(x,y,z)$ and zenith and azimuth
angles) or in spherical coordinates (vertex ${\bf R}(r,\theta, \phi)~
)$ for a track candidate with at least 6 PMTs hit.


\section{Pattern Recognition}


\subsection{$^{40}\!K$ filter and hit selection}
One of the problems with a sea $\nu$-telescope is $^{40}K$ noise.
The first part of chameleon needs to clean up as much noise as
possible, since a \cs re\-con\-stru\-ction is very sensitive to noise. 

To use the potential of multi-PMT OMs, the algorithm uses a slightly
modified  wrt the proposed ``write up string designs'' \cite{elswriteup} scheme. 

The first
modification is that for each PMT {\em only} the first photon is
considered to be a hit, while the charge is set equal to the total
number of photons this particular PMT registered  (i.e. one has only
one hit per PMT). This was deemed necessary until the final time over
threshold mechanism in the MC is stabilized. 
The second modification is that not only adjacent PMTs are taken into
account (see below).

The $^{40}\!K$ filter is based on a  two-layered algorithm, the trigger
algorithm and the hit selection.

\begin{figure}[!htpb]%
  \centering
  \subfloat[Time differences between succesive hits per OM.]
  {\includegraphics[width=90mm]{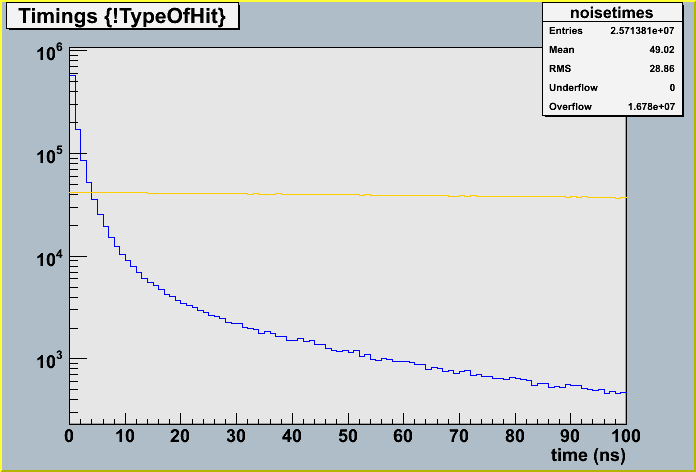}}$\;$
  \subfloat[Ratios of \v{C}erenkov photons and noise photons  wrt the
  total number, for the same
  sample. ]
  {\includegraphics[width=90mm]{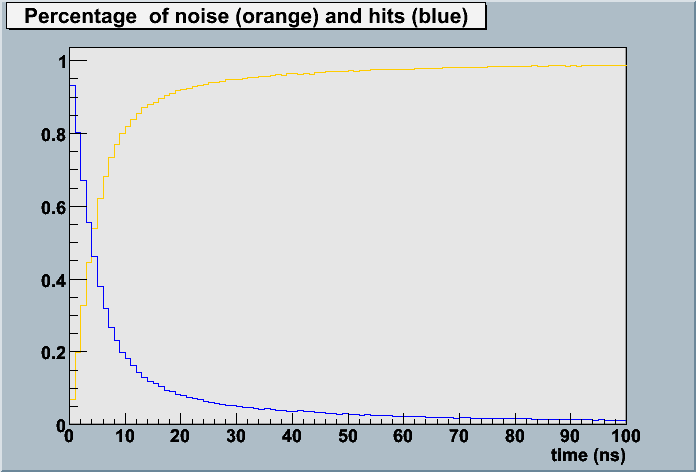}}
 
 \caption{Time differences between pairs of succesive hits per OM.
   Blue = both hits are photons, orange = at least one hit is noise.}
 \label{fig:timediffs}
\end{figure}

{\em The trigger
algorithm} is based 
roughly on the L1 trigger (see  \cite{elswriteup}), as broadened after
a MC study of timings for photons.
Specifically, for all the hits that are registered on an OM, the first
thing to do is to select
the data set that will be sent to the reconstruction. Since signal
photons are concentrated around the track that created them while
noise is randomly produced, a first filter can be based on time
coincidences of hits. In fig.(\ref{fig:timediffs}) the differences in
time of arrival between succesive hits for OMs with 2 or more hits are plotted 
for noise production rate of 8 kHz per PMT. It should
be noticed here that although the expected noise level for the
Multi-PMT design is at 5.2 kHz per PMT, a higher level of noise was
used for reasons, in order to be on the safe side.  If both hits are
photons, they are registered in blue. It can be seen clearly that when
two succesive hits on the same OM are registered within 6 ns or less,
then they are most probably both real photons.


\begin{figure}[!htpb]
  \centering
\includegraphics[width=100mm]{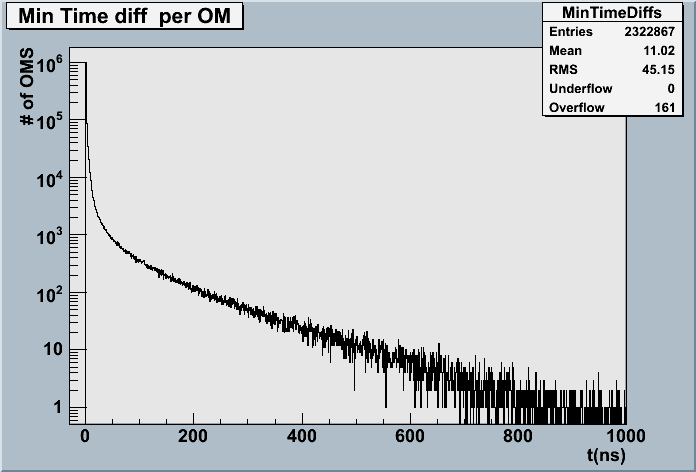}  
  \caption{For each OM with more than  one photon, 
    choose the 2 photons which are closest in time and plot their time
    difference. One such pair per OM can be plotted. }
  \label{fig:mintimediffs}
\end{figure} 


\begin{figure}[!htpb]%
  \centering
  \subfloat[Time residual for hits with charge equal to  1.]
  {\includegraphics[width=100mm]{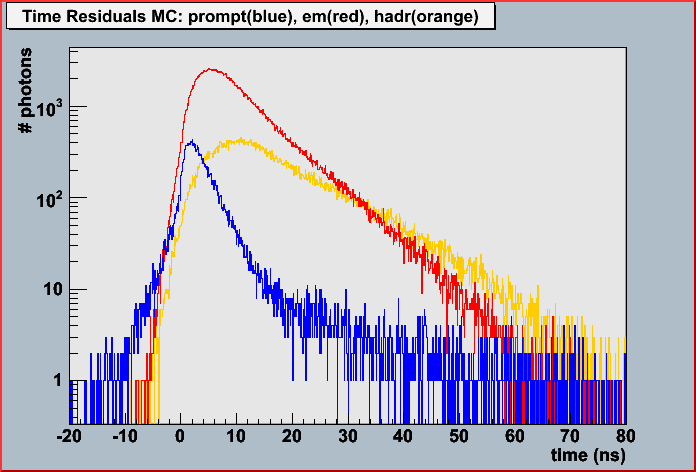}}\qquad
    \subfloat[Time residual for hits with charge equal to 3.]{\label{3figs-c}
\includegraphics[width=100mm]{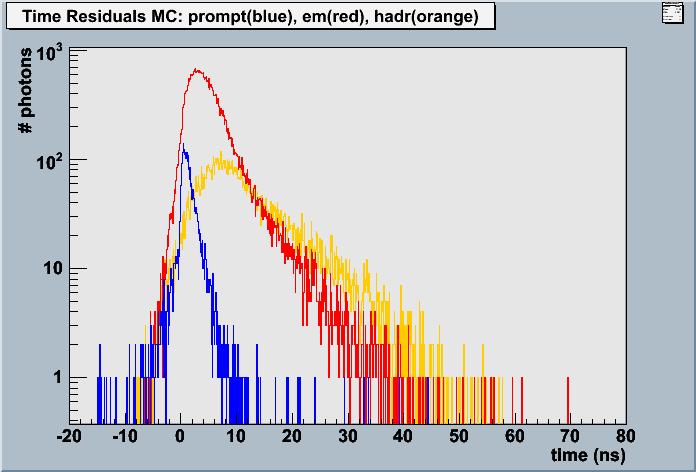}}%
  \caption{Time residuals.}
  \label{fig:2figs}
\end{figure}

The detailed description of these algorithms  is as follows:

\begin{itemize}
\item Only use the {\em first} photon for each PMT within an
  event. This is done in order to keep the results totally unbiased as
  to the exact electronics of the system. After the finalization of
  the T/threshold mechanism, the information from the rest of the
  photons will be used.

\item Start by looping over all the OMs of the event. At
  this layer consider only OMs with 2 or more detected hits.
  \begin{itemize} 
  \item After ordering the hits of each OM in time, find which
    two of them are closest. Check their time difference $\Delta t$: If
    the following condition  does not hold
    \begin{equation} 
      \label{eq:loc}
      |\Delta t|< 6 \; ns,
    \end{equation}
    the whole OM is
    discarded, as it is assumed that since the smallest time
    difference is larger than the OM's diameter the photons are too
    sparse to be real photons, see e.g. \cite[p.60]{hart} and fig. (\ref{fig:timediffs}). 
  \item In case eq. (\ref{eq:loc}) is satisfied, then perform a local
    loop over the  photons of this particular OM, starting from the
    minimum time pair.  Keep {\em only}
    those photons that are no further away than  $7\;ns$ from the
    previous photon.
  \end{itemize}
\item After this first filtering, find those OMs that have registered
  the largest numbers of hits (number of PMTs times charge).
  Only keep the event if the ``largest'' OM has registered hits in at
  least  3 of its PMTs.  These ``large'' OMs are subsequently grouped 
  by causal connection. Each of these clusters of ``large'' OMs will be used as a
  basis for track reconstruction in the next steps, leading possibly
  to multiple tracks.
\item Loop over these  OM's closest neighbours, i.e. those OMs
  that are located within 180 m ($\sim $ 3 absorbtion lengths) from
  the OM with the largest hit. The closest neighbors are OMs that did
  not pass the trigger, so by checking causal connections with the
  large OMs, more hits can be retrieved.
  For string distances larger than 180 m this means that this search
  is reduced to same-string OMs.
\item Obviously, the number of active (i.e. that have at least one hit on one of their 
PMTs) neighbouring  OMs is smaller than the number of neighbouring OMs.
The photons on the closest neighbours' PMTs are required to obey
  the condition 
\begin{equation} 
    -T_0 <  |\Delta t|-\frac{r}{c} < t_0,
\label{eq:timediff}
    \end{equation}
 where $r$ is the distance between OMs under consideration, $T_0$ and $ t_0$ are arbitrary
 constants and the time difference 
$\Delta t$ is
 calculated between the current photon's time and the time of the
 first photon of the ``large'' OM under consideration (For a justification see e.g. \cite[fig 5.2]{kuch}.)


In fig. \ref{fig:causal},
the quantity $ |\Delta t|-\frac{r}{c}$ of eq. (\ref{eq:timediff}) is
plotted for a  
data sample of a small  detector with string distances 
at only 90m. 
As can be seen from the first of these figures, 
for a neighbourhood of $R=120$ m around the largest OM, 
one can take into account all the photons below $t_0$, since there 
the $^{40}\!K$s remain a small percentage of the total number of photons.
Unfortunately, as the radius increases, the local time window, equal
to 
the time it takes a muon to travel one such radius $\sim R/c$,  
is large enough for more $^{40}\!K$ photons to be considered:
$R/c = 400, 600, 667$ ns, for the 3 radii. In the following tests
$T_0$ was set 
to 300 ns for $R=180$ and $200$, while it is not important for 
120 m where the filter is more effective (in the latter case of 
course there are fewer photons so that the next step of reconstruction is harder).

\begin{figure}[!htpb]
  \centering
\subfloat[Photons in a sphere of 120 m.]
 {\includegraphics[width=90mm]{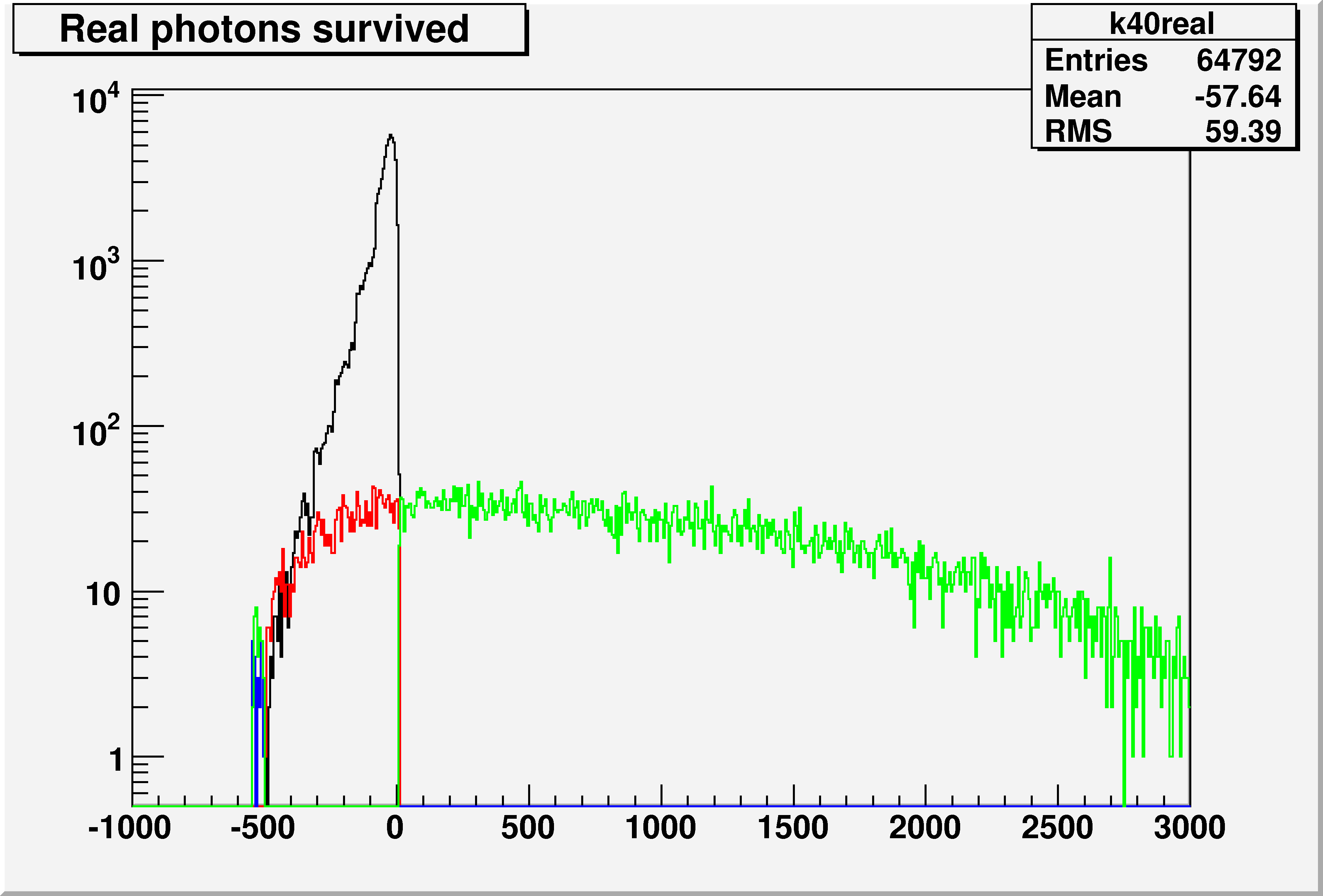}}
\subfloat[Photons in a sphere of 180 m.]
 {\includegraphics[width=90mm]{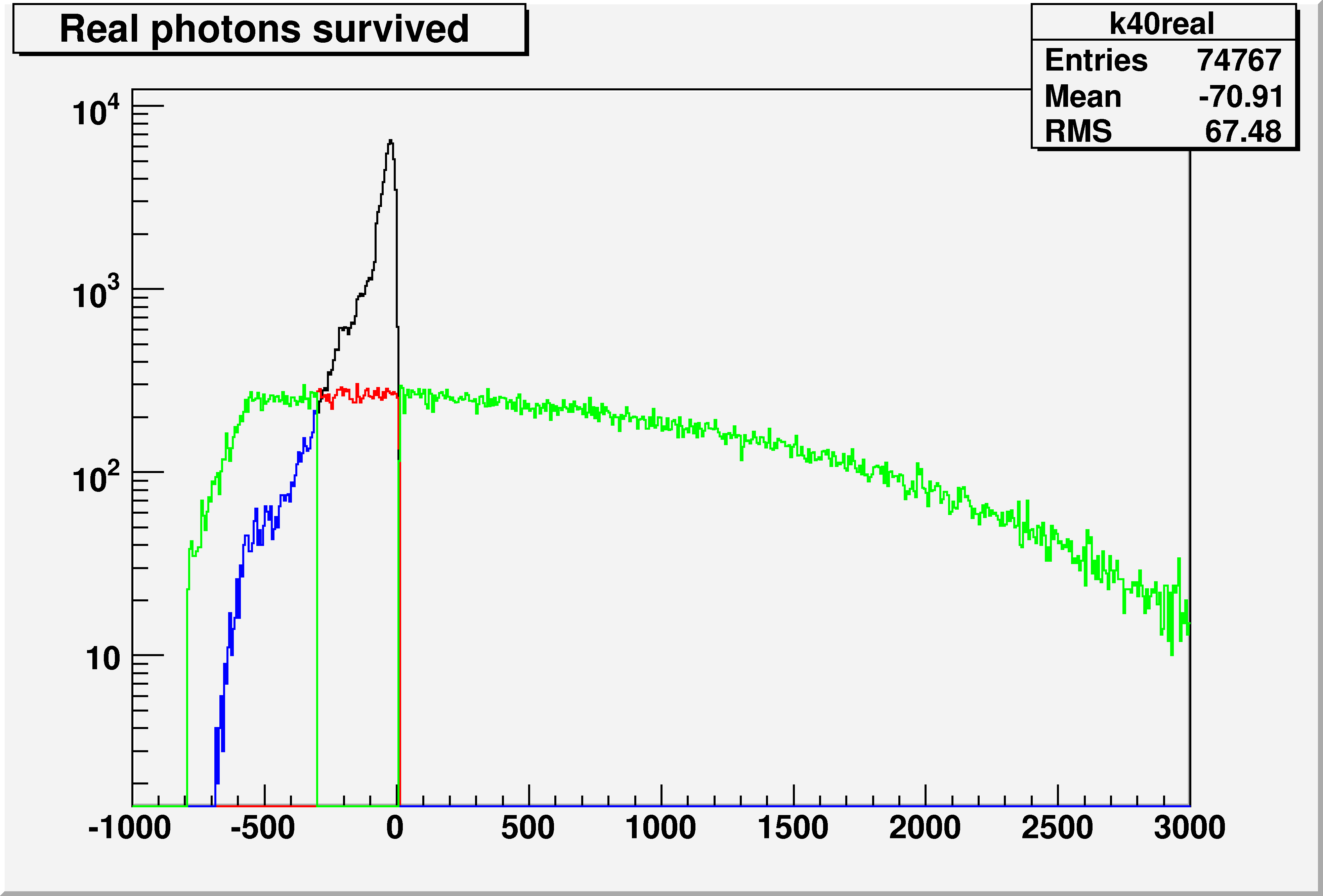}}\\
\subfloat[Photons in a sphere of  200 m.]
 {\includegraphics[width=90mm]{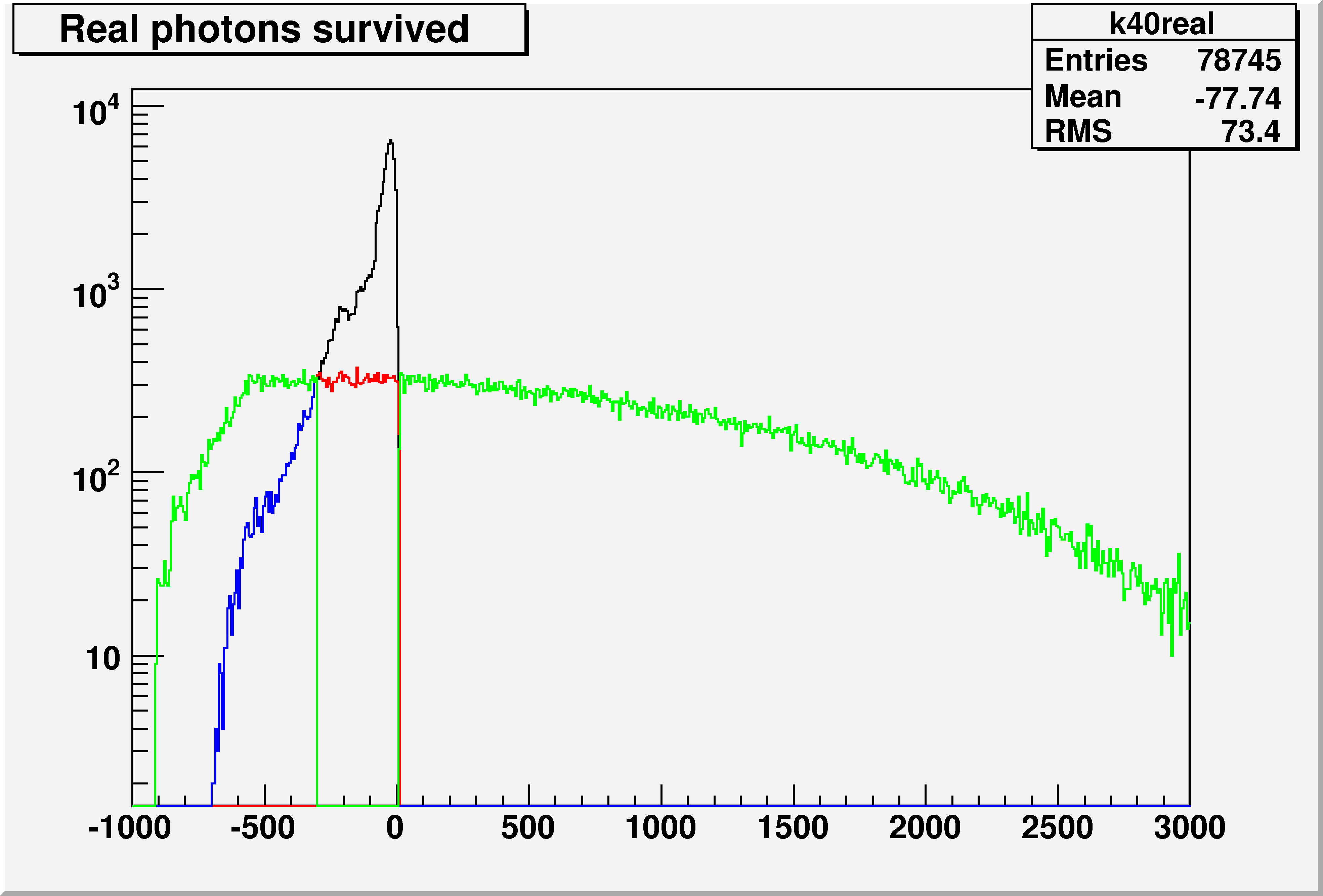}}
  \caption{A plot of eq. (\ref{eq:timediff}). Only photons within a
    sphere of 120, 180 and 200 m 
    respectively are taken into account. Black: survived real photons. Red: Survived $^{40}\!K$. 
    Blue: Rejected real photons. Green: Rejected  $^{40}\!K$. }
  \label{fig:causal}
\end{figure}

The combination of these 2 filters eliminates most of   $^{40}\!K$
photons, as can be seen in fig. \ref{fig:local}. A major problem whcih should be considered 
is the fact that
 as the number of OMs under consideration rises, so does the relative number of $^{40}\!K$
 hits which survive the causality filter.
\begin{figure}[!htpb]
  \centering
\includegraphics[width=90mm]{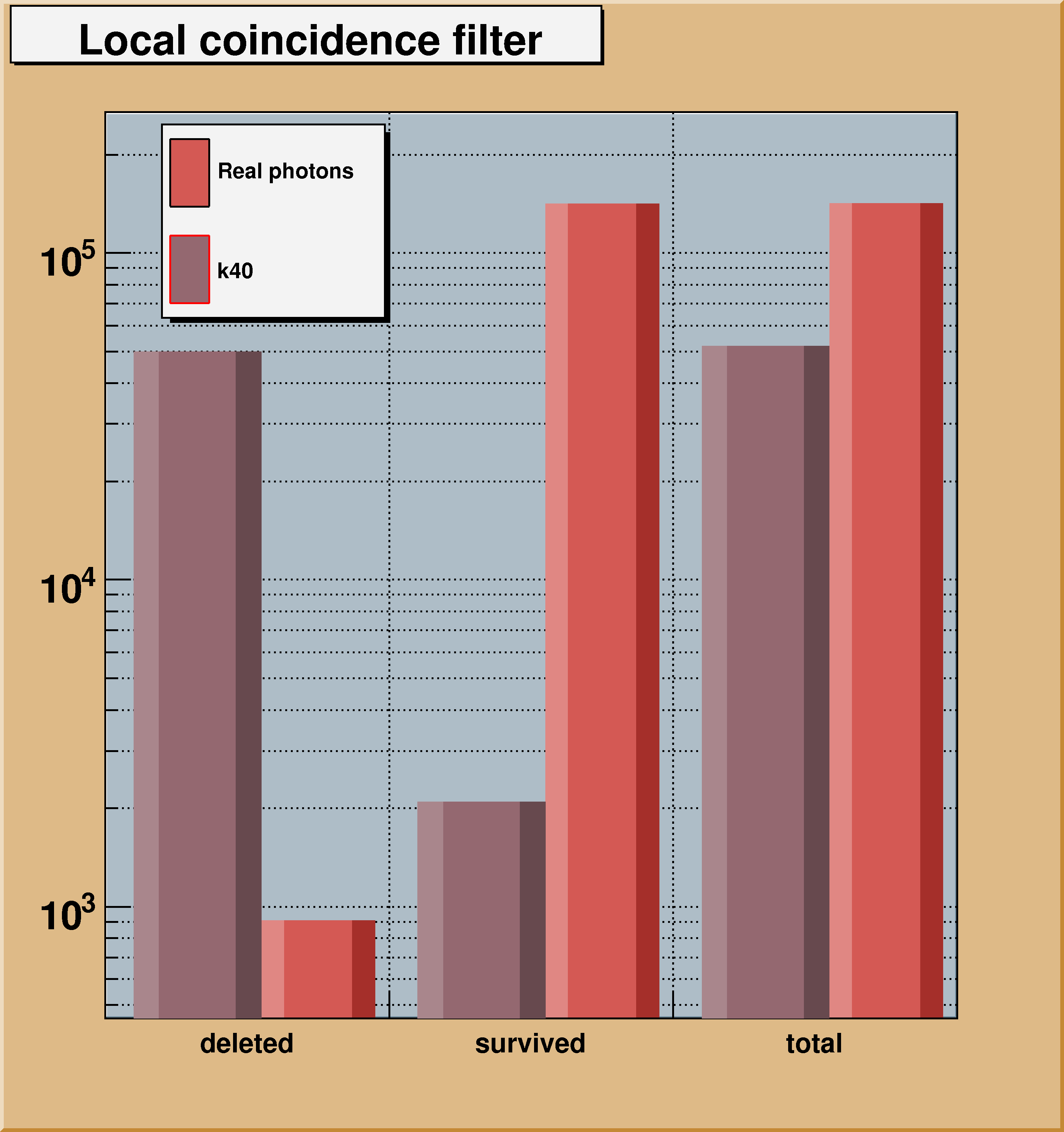}  
  \caption{The results of the $^{40}\!K$ filtering-out process over all
   OMs with 3 or more hits plus the largest hit of each event and its
   closest neighbours. String distance  120 m.}
  \label{fig:local}
\end{figure} 
\item What we are left with at this stage is a series of clusters of
  ``large'' OMs, together with isolated hits from their neighbouring
  OMs, which are all causally connected. Every such group of causally connected OMs is
  then passed through a {\em fit with rejection}  method (see \S
  \ref{sec:fit}) with a $N$ parameter equal
    to $5~\sigma$.  The resulting
    track is used to search again for isolated hits. The search is performed on all hits that were not
    used in the previous steps and which are within $~2$ absorption
    lengths from the track, on the basis of the
    time
    residuals of these hits wrt the track. After the search, and if
    additional hits were found,  a  final {\em fit with
      rejection} at $5~\sigma$ is performed again.
    
\end{itemize}


\subsection{Track Parameter Estimation}

One of the main problems with track reconstruction is the fact that
(depending on the track energy) most of the photons produced by it do
not come directly from the muon, but are produced through
brehmsstralung processes {\em away} from it. The result is that the
errors ascossiated with the reconstruction are not expected to be
gaussian in form, or equivalently, the \cs-probability distribution is
not expected to be flat. The decision was taken to always pass the
reconstructed track from a final constant deweight minimization, which
aggravates the \cs-probability distribution (by assigning a \cs
probability close to 1 or 0 for most of them), in order to obtain a
more representative pull distribution for the track parameter errors.
For details, see \S \ref{sec:recoperf}.

\subsection{Final Track Selection}
As already noted, the above procedure allows for multiple tracks to be
reconstructed. In order to eliminate ghost tracks, a search is
performed  amongst the reconstructed tracks for common hits. In case such common hits are
identified, the track with the smaller overall number of hits is
deleted if more than 10\% of its hits were shared  with the track with
the larger
number of hits.

\section{The User Point of View}
Thanks to the modular construction of SeaTray, the user only has to
decide about the basic parameters of the reconstruction and implement
them in the driving chameleon module in a simple python script.

The parameters are:
\begin{itemize}
\item The names of input and output files (both {\tt *.i3} and {\tt
    *.root} files are supported for the output)
\item The names of the various containers for particles and photons
  (input MC particles, output reco particles, hits, pulses etc.)  
\item The  ``{\tt HitsWeightingMethod}'' parameter, which specifies
  the way hits used are going to be reweighted during reconstruction.
  This parameter accepts one of three possible values:
  \begin{itemize}
  \item ``{\tt constant}'':   This is the default. In this case 
    the error is set by hand in the parameter ``{\tt deweightval}''
    (see below) for all the hits of the track .
  \item ``{\tt residuals}'': The error is set by the relative time residuals of the hit vs the recotrack.
  \item ``{\tt deweight}'':  The deweight and fit method, a recursive
    method which calculates the hit residuals and deweights those hits
    that are $N\sigma$ away from the nominal value of $2 \; ns$. $N$
    is set by the  ``{\tt deweightval}'' parameter (see below).
  \end{itemize}
\item The  ``{\tt deweightval}'' parameter, which is used with the
  ``{\tt constant}'' and ``{\tt deweight}'' fitting methods above.
\end{itemize}

The  ``{\tt deweightval}'' and ``{\tt residuals}'' methods were
developped for the study of the MC data, but were not used in the
results shown in this report.

\subsection{ Analysis}

The analysis of the output data can be performed with the help of a
series of modules that export the relevant quantities in chains of root files
(see directory {\tt AnalysisUtility}). 
A series of scripts were written in order to make plots etc (within
the  {\tt scripts} directory). 

The implementation of the {\tt boost} libraries within SeaTray allows for 
the use of both {\tt ROOT} and python (i.e.. {\tt pyROOT}) scripts for the same
task (see e.g. in the {\tt scripts} directory files {\tt timeDiffspyROOT.py}
and {\tt analysis/src/analysis\_tree.C}).

In cases where there is need for more computational speed, a relevant
C++ class can be found in {\tt AnalysisUtility}, which is also implemented
in a way that allows for its usage from within both python and  CINT
(ROOT's C++ interpreter). For an example see  {\tt EffAreaPlotsMaker} and its
 C++ avatar {\tt useMakeEffAreaPlotsClass()}, in
{\tt analysis\-/src\-/analysis\_tree.C} and in pythonic form in {\tt angularPlotspyROOT.py}.
 

\chapter{Reconstruction Performance}

\section{Reconstruction Performance}
\label{sec:recoperf}

\begin{figure}[htp]
  \centering
  \centering
  \subfloat[The efficiency of the reconstruction vs the energy  of CC
    muons]
  {\label{fig:muentot}
    \includegraphics[width=75mm]{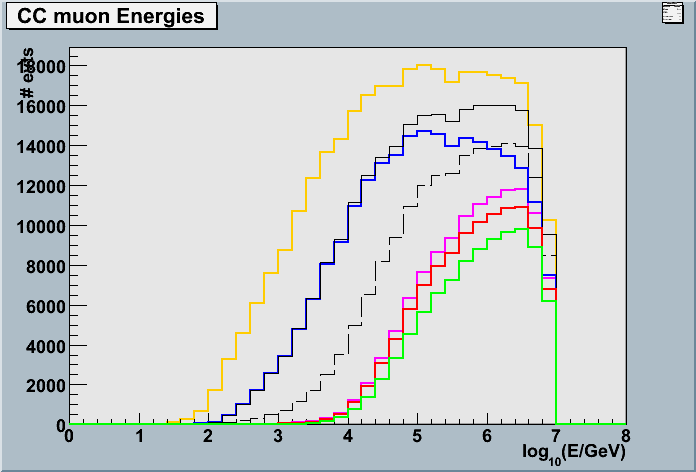}}\quad
   \subfloat[Normalized efficiency of the reconstruction. Divided by
   the number of  generated  CC muons with more than 6 OMs hit in the
    sample. The fluctuation for $E<100 GeV$ is due to
    small statistics. ]
  {\label{fig:muennorm}
    \includegraphics[width=75mm]{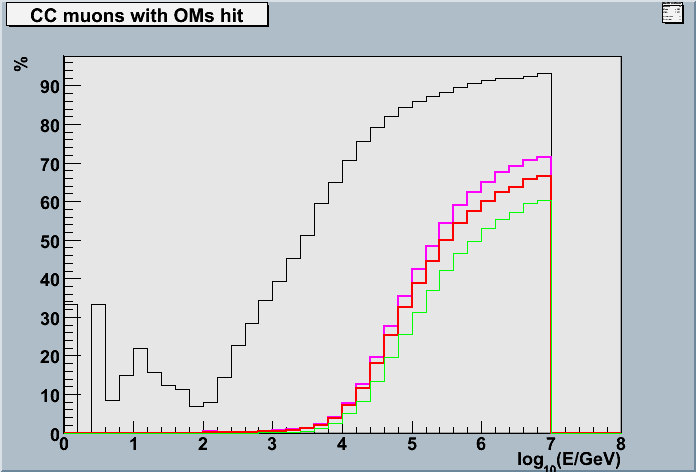}}
  \caption{The efficiency of the reconstruction vs the energy  of CC
    muons. 
    {\it Orange line:}  Generated CC muons with more than 6 OMs hit in the
    sample. 
    {\it Black line:} Generated CC muons with more than 10 OMs hit in the
    sample. 
    {\it Blue line:} Contained generated CC muons with more than 10 OMs hit. 
    {\it Dashed Black line:} Reconstructed muons with more than 6
    OMs in the track. 
    {\it Magenta  line:} Reconstructed muons with more than 10 OMs in the track.  
    {\it Red line:} Reconstructed muons with more than 10 OMs in the track,
    reconstructed within
    less than $5^{\circ}$ of the MC direction. 
    {\it Green line:} Reconstructed muons with more than 10 OMs in the track, reconstructed within
    less than $0.4^{\circ}$ of the MC direction. 
    } 
  \label{fig:muen}
\end{figure}

The efficiency of the reconstruction algorithm is shown in
fig. \ref{fig:muen}. The orange line represents the energy spectrum of
all MC muons that gave hits on at least 6 OMs (\v Cerenkov photons
originating from the muon). That is the theoretical
minimum of needed OMs for a reconstruction.

 This spectrum is compared
to the spectrum of reconstructed muons (e.g. the green line
represents those tracks which contain at least 10 OMs and their
direction is within $0.4^{\circ}$ of the MC direction). 
Out of a sample of 5,800,000 neutrinos,  492,147 were
CC events with 10 or more hit OMs, 253,386  of which were
contained\footnote{We define a track to be contained if it passes
  through the detector, i.e. if the line of the track crosses at least
one of the 8 planes that define geometrically the detector.}. 
There were  122,302  reconstructed  events with 10 or more  OMs in the
track, 118,930 of which originated from a CC muon. 

\begin{figure}[htp]%
  \centering
  \subfloat[Reconstructed muons with 6 OMs.]
  {\label{fig:totatfakes}
    \includegraphics[width=75mm]{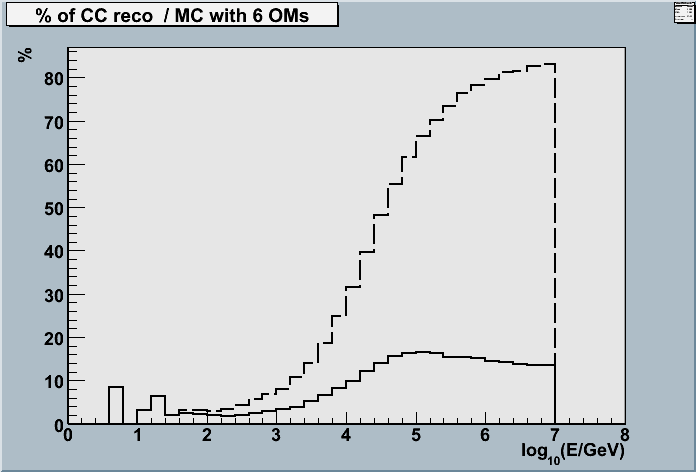}}\quad
  \subfloat[Reconstructed muons with 8 OMs.]
  {\label{fig:8fakes}
    \includegraphics[width=75mm]{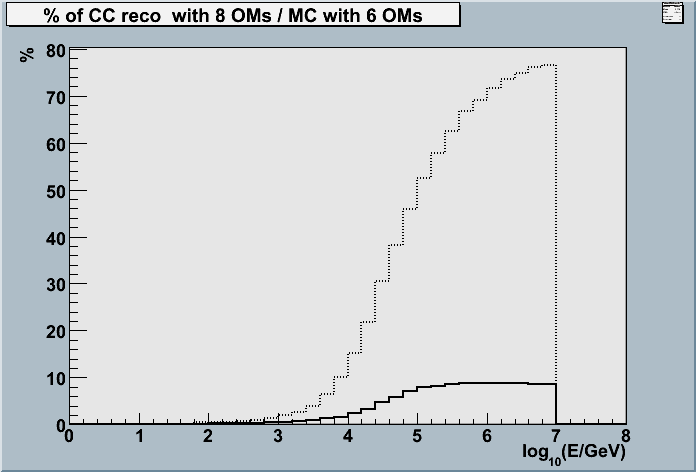}}\\
   \subfloat[Reconstructed muons with 10 OMS.]
  {\label{fig:10fakes}
    \includegraphics[width=75mm]{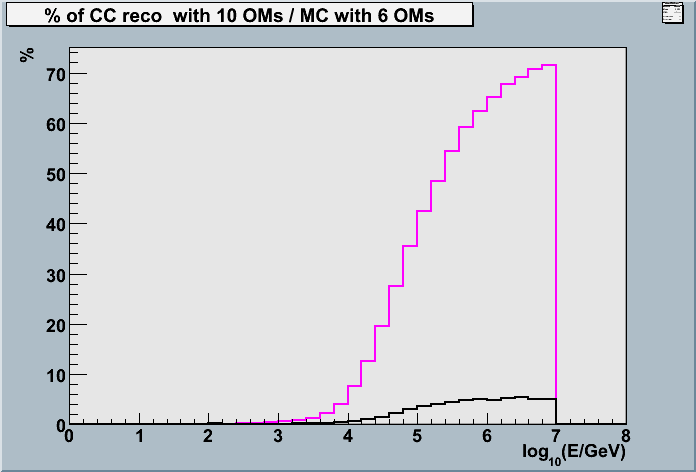}}\quad
   \subfloat[Reconstructed muons with 12 OMS.]
  {\label{fig:12fakes}
    \includegraphics[width=75mm]{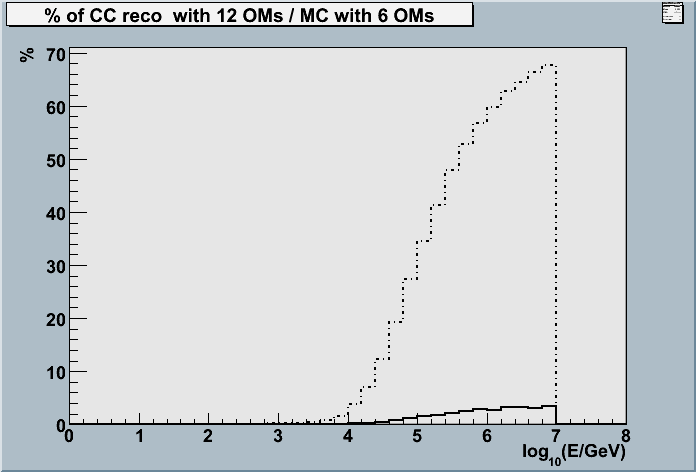}}\\
  \caption{The efficiency of the reconstruction vs the energy:  Reconstructed tracks after cuts, 
    compared with the respective numbers  of 
    ``fakes'', i.e. reconstructed muons  with a reconstructed
    direction {\em more}  than $5^{\circ}$ off of
    the MC muon (solid black lines). All plots  normalized to the number of Generated CC
    muons with  6 or more OMs hit in the sample.} 
  \label{fig:fakes}
\end{figure}
\begin{figure}[htpb]
  \centering
  \subfloat[$\theta$ difference.]
  {\label{fig:thetadiff}
    \includegraphics[width=80mm, height=120mm]{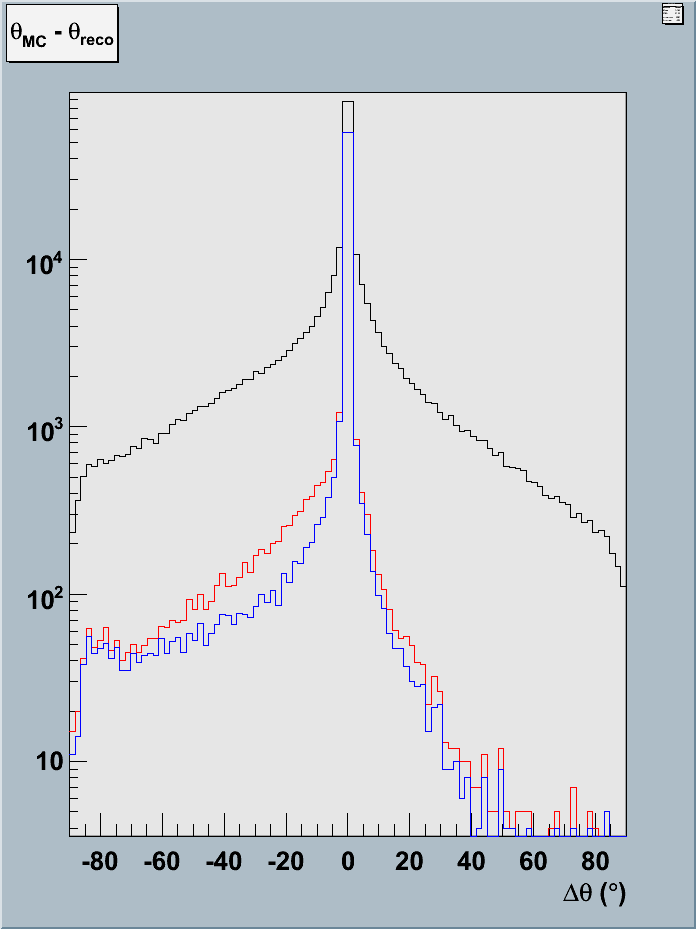}}
  \subfloat[$\phi$ difference.]
  {\label{fig:phidiff}
    \includegraphics[width=80mm, height=120mm]{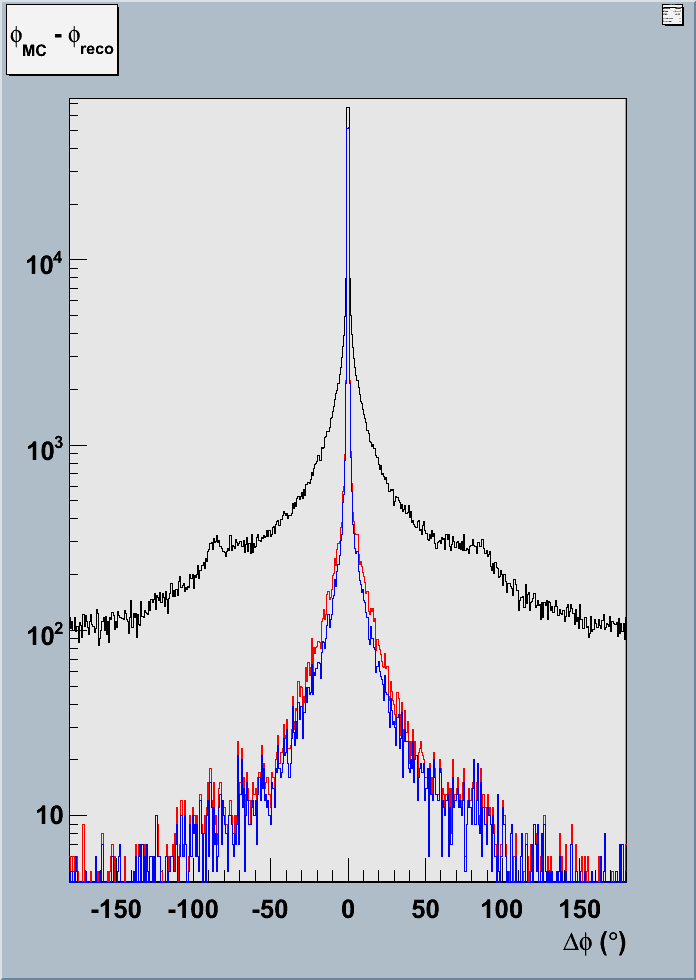}}
  \caption{Differences in angles  between MC muon and reconstructed
    muon in deg.  Black line: all reconstructed muons, Red line: muons
    with at least  10 OMs, Blue line: CC only events with at least 10 OMs. }
  \label{fig:anglediff}
\end{figure}

Additionaly, fakes are shown in fig. \ref{fig:fakes}. By ``fakes'' here
we designate those reconstructed tracks whose direction is  $5^{\circ}$
or more away from the MC direction. 
The differences in reconstructed and MC $\theta$ and $\phi$ angles are shown in fig. \ref{fig:anglediff}.

\begin{figure}[htpb]
  \centering
\subfloat[Before cuts.]
{  \label{fig:trackangle}
  \includegraphics[width=120mm]{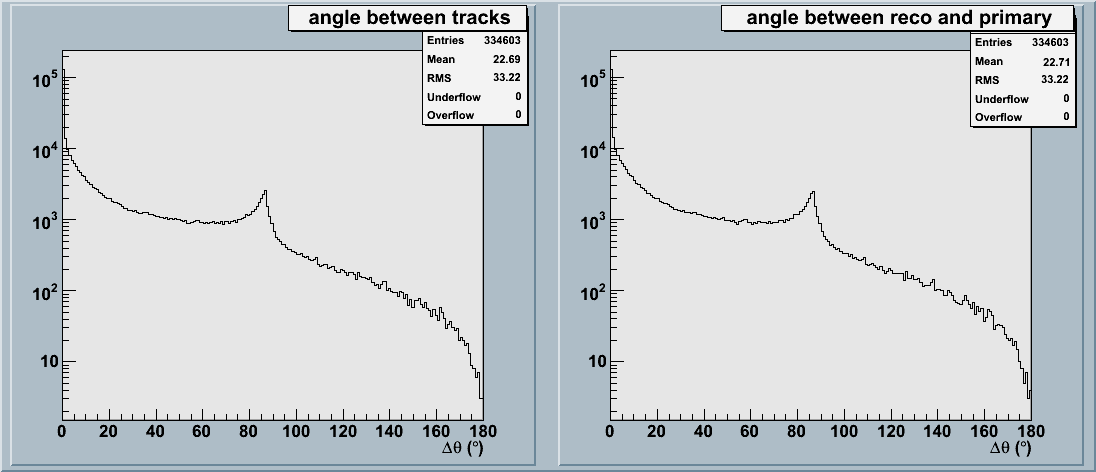}}\qquad
\subfloat[After cuts (10 OMs in reco track).]
{  \label{fig:trackangle_cut}
 \includegraphics[width=120mm]{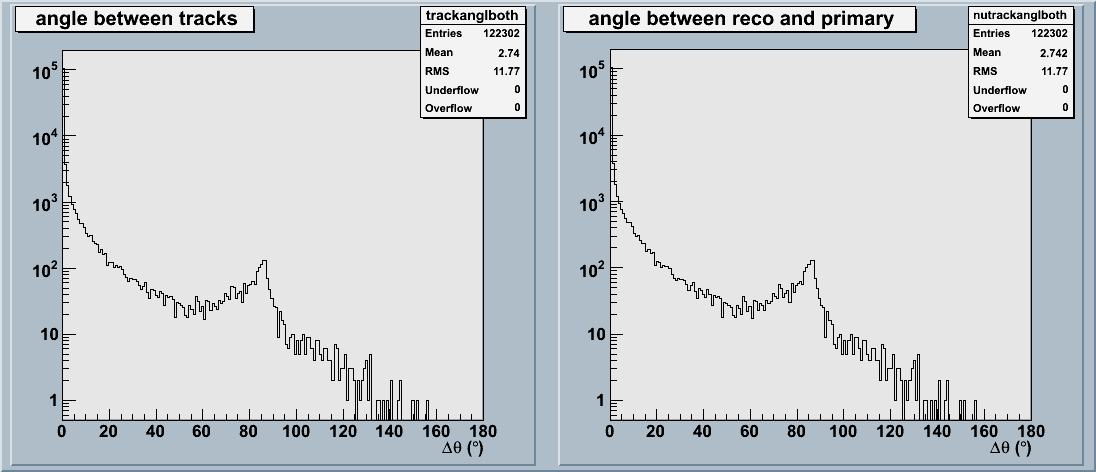}}
  \caption{Angle between reconstructed track and muon (left), track
    and neutrino (right).}
\label{fig:angle}
\end{figure}

\begin{figure}[htpb]
  \centering
  \includegraphics[width=80mm]{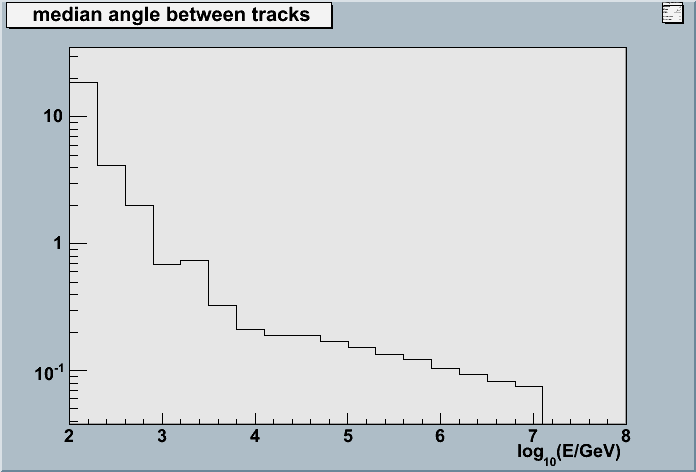}
  \caption{Median between reconstructed and MC muon as a function of $\log_{10}(E)$.}
  \label{fig:diffmedian}
\end{figure}

 The cut of 10 OMs for the acceptance of a track
as well reconstructed follows from  plots  \ref{fig:fakes} 
 since it is shown there that a resonable compromise
between the reconstruction efficiency and fakes is achieved for this cut.
This is further corroborated by fig. \ref{fig:angle}, where the
difference in  the reconstructed and the MC neutrinos and muons is
shown, before and after the cut. The overall median angle for these
plots is:
\begin{itemize}
\item Between primary and reconstructed, before cuts, $4.55^{\circ}$.
\item  Between primary and reconstructed, after cut, $0.12^{\circ}$.
\item Between MC muon and reconstructed, before cuts, $4.53^{\circ}$, and
\item between MC muon and reconstructed,after cut, $0.11^{\circ}$.
\end{itemize}
The differential median angle between MC muon and reconstructed per
energy bin is shown in fig. \ref{fig:diffmedian}.

\subsection{ A note on the asymmetry of  fig.  \ref{fig:thetadiff} and
on ghost solutions}
\label{sub:asymmetry}
\begin{figure}[htpb]
  \centering
  \subfloat[$\theta$ difference, downgoing muons  $ z\in ( 0^{\circ},90^{\circ} )$.]
 {\label{fig:upcoming}
  \includegraphics[width=80mm]{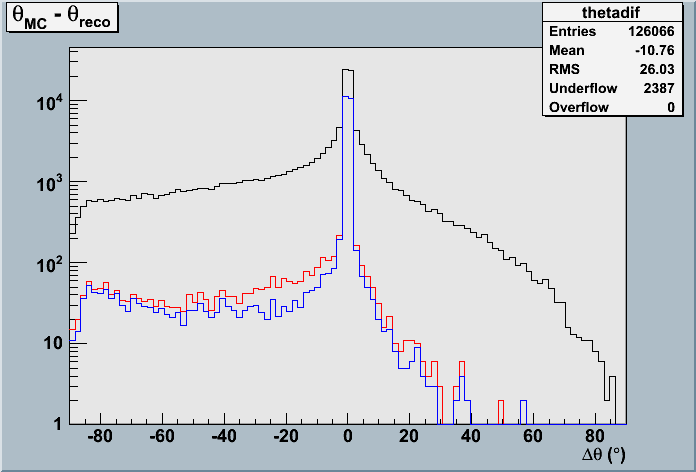}}
  \subfloat[$\theta$ difference,upcoming muons
  $z\in ( 90^{\circ},180^{\circ} ) $.]
 {\label{fig:downgoing}
\includegraphics[width=80mm]{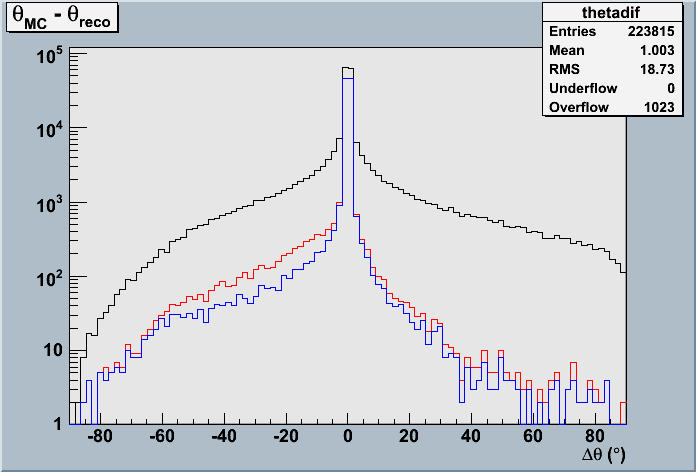}}
  \caption{$\theta$ difference, downgoing and  upcoming. Color code:
    see fig. \ref{fig:anglediff}  }
  \label{fig:theta2}
\end{figure}

\begin{figure}[htpb]
  \centering
  \includegraphics[width=80mm]{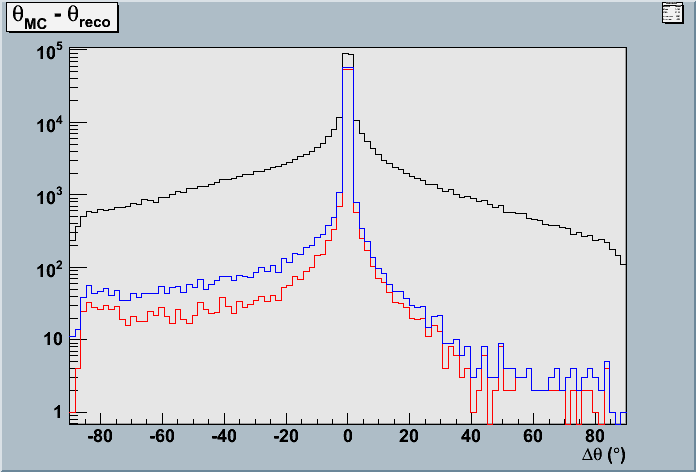}
  \caption{Demonstration of the up-down asymmetry (see \S \ref{sub:asymmetry}). Black: MC data. Red:
    Reconstruction, 10 OMs.
    Up: Declinations.}
  \label{fig:no_hadr}
\end{figure}

There is an  obvious asymmetry in fig.  \ref{fig:thetadiff} between up
and down. This is shown
separately  for upcoming (fig.
\ref{fig:upcoming}), and downgoing (fig. \ref{fig:downgoing}) tracks.
This asymmetry can be possibly traced for its most part to hadronic processes,
which emit undirectional spherical waves of photons. As a
demonstration, in fig. \ref{fig:no_hadr}, the $\theta$ differences are
plotted again, but for the red line (tracks with at least 10 OMs
included), we did not include those tracks that more than half of
their hits are of hadronic origin. Comparison with fig.
\ref{fig:thetadiff} shows immediately that the number of
misreconstructed tracks has dropped by almost one order of magnitude.

 The physics of the asymmetry might possibly
be attributed to the detector asymmetry: OMs are not spherical
symmetric (there is no top PMT). Since then  OMs are slightly up-down
asymmetric  (one more PMT in the
south hemisphere), the tracks' $\theta$ angles (we remind that $\theta
= 0$ corresponds to a track coming from the nadir) tend to be slightly
pulled towards the zenith when symmetric light is emmited (as is the
case for hadronic processes): the OMs accept more light on their south hemispheres
from a spherical wave than on their north hemispheres, thus
interpreting the spherical wave as an upcoming track.
  Work is in progress to implement a hadronic processes
module, able to identify and reconstruct correctly these tracks.

\begin{figure}[htpb]
  \centering
  \includegraphics[width=80mm]{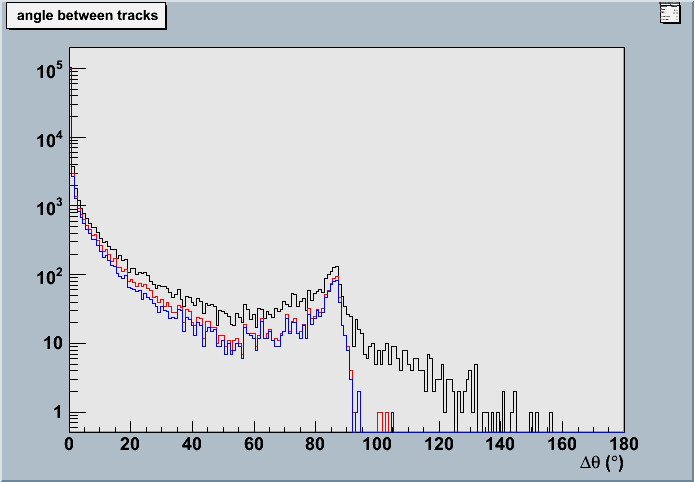}
  \caption{Angle between reconstructed track with at least 10 OMs hit
    and muon (black). The same for tracks with less than half of their
  hits of hadronic origin (red).}
  \label{fig:trackanglenohadr}
\end{figure}

The problem of ghost solutions is of different origin. As it can
be seen in fig. \ref{fig:trackanglenohadr}, where the same artificial
cut on hits of hadronic origin was applied, although the tails are
minimized after the cut, the kink at around $90^{\circ}$ remains. The
percentage of the misreconstructed tracks around the kink (from
$60^{\circ}$ to $100^{\circ}$) is  1.04\% of the total reconstructed
tracks with at least 10 OMs.


\subsection{ Quality of the reconstruction}
\label{sub:pulls}
\begin{figure}[htpb]
  \centering
  \includegraphics[width=80mm]{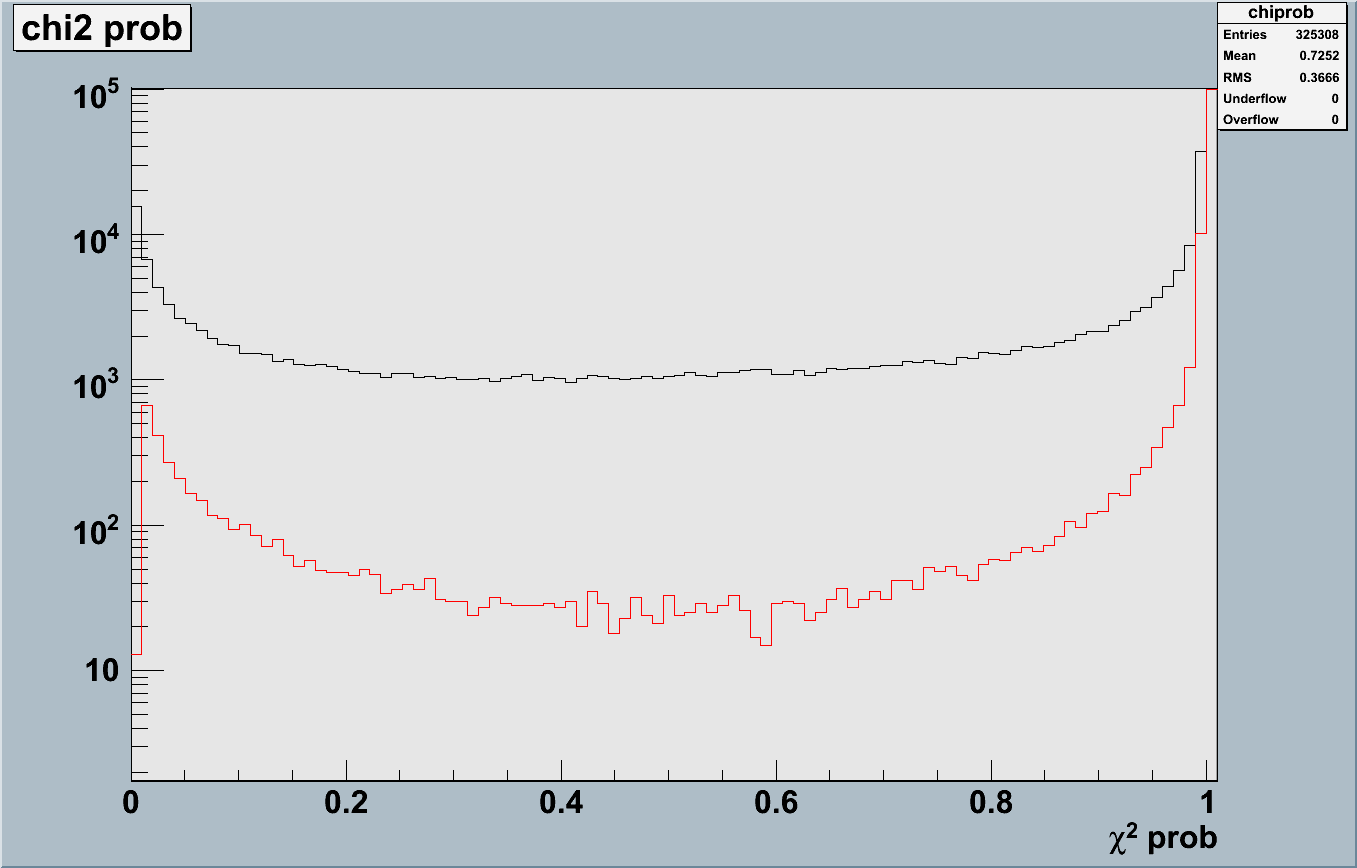}
  \caption{$\chi^2$ prob distribution. Black: All tracks. Red:
    reconstructed with 10 or more OMs.}
  \label{fig:chiprob}
\end{figure}

\begin{figure}[htpb]
  \centering
\subfloat[$\phi$ pulls
($\displaystyle\frac{\Delta\phi}{err}$).]
{  \label{fig:phipul}
  \includegraphics[width=120mm]{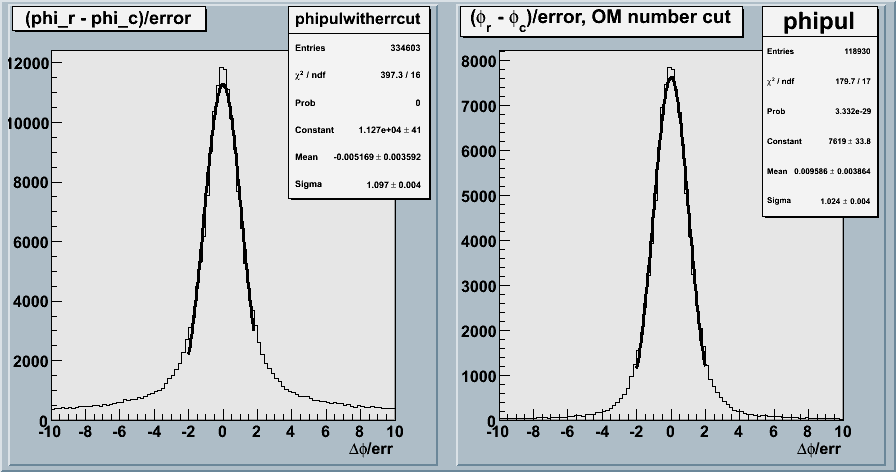}}\qquad
\subfloat[$\theta$ pulls 
($\displaystyle\frac{\Delta\theta}{err}$). ]
{  \label{fig:thetapul}
 \includegraphics[width=120mm]{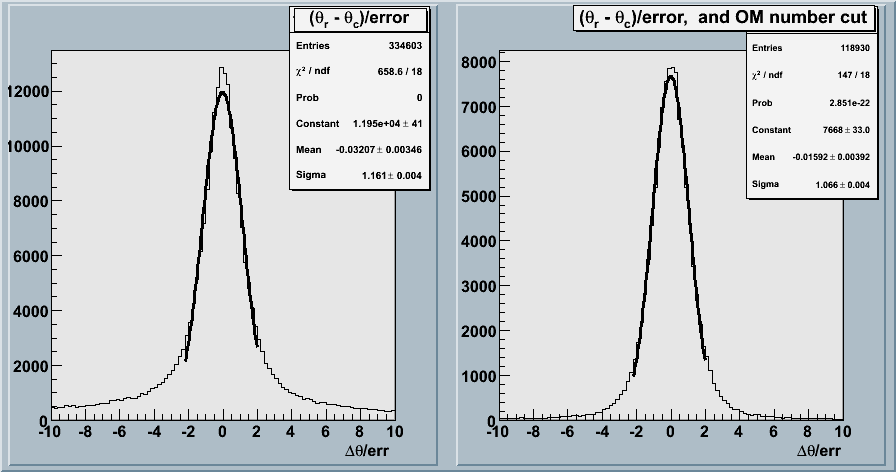}}
  \caption{Pulls for $\phi,  \theta$. Left: all tracks, Right: tracks
    with 10 OMs}
\label{fig:pul}
\end{figure}

\begin{figure}[htpb]
  \centering
  \includegraphics[width=100mm]{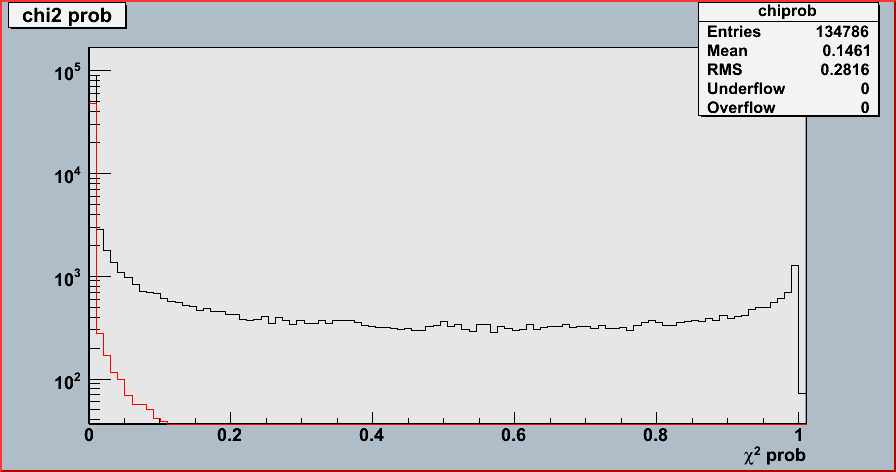}
  \caption{$\chi^2$ prob distribution for a constant deweight fitting
    with error = 2 (see text, \S. \ref{sub:pulls}). Black: All tracks. Red:
    reconstructed with 10 or more OMs.}
  \label{fig:chiprob2}
\end{figure}

\begin{figure}[htpb]
  \centering
\subfloat[$\phi$ pulls ($\displaystyle\frac{\Delta\phi}{\sigma}$).]
{  \label{fig:phipul2}
  \includegraphics[width=120mm]{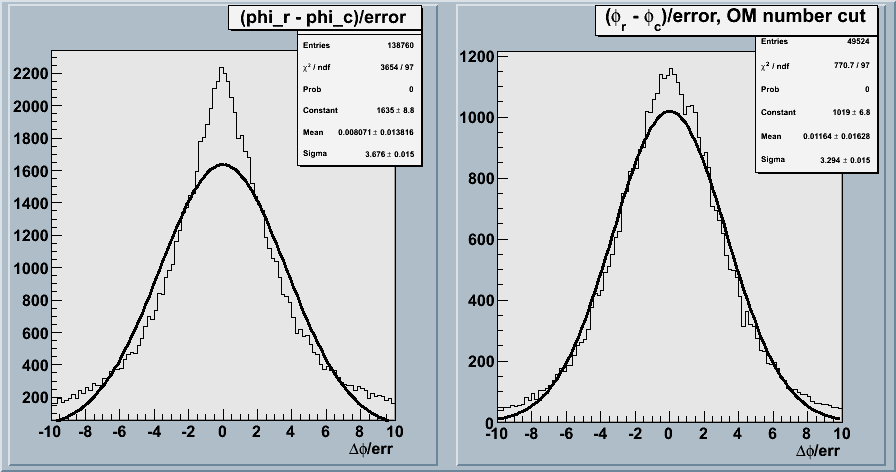}}\qquad
\subfloat[$\theta$ pulls ($\displaystyle\frac{\Delta\theta}{\sigma}$).]
{  \label{fig:thetapul2}
 \includegraphics[width=120mm]{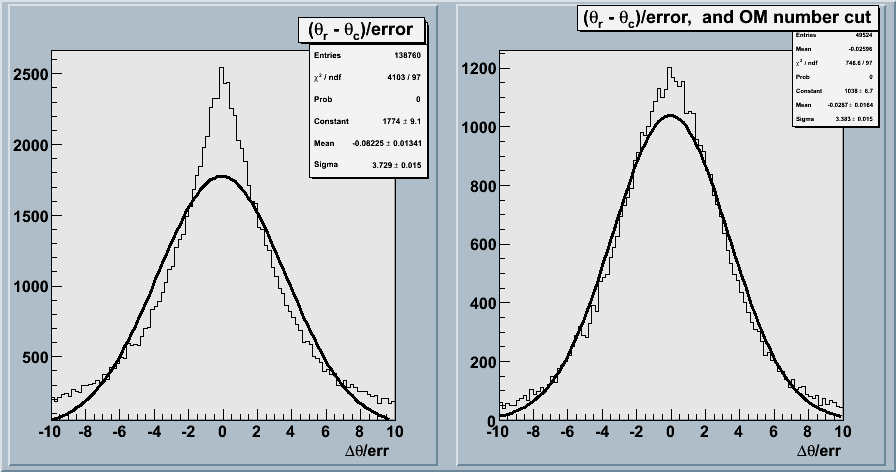}}
  \caption{Pulls for $\theta, \phi$ for a constant deweight fitting
    with error = 2 (see text, \S. \ref{sub:pulls}). Left: all tracks, Right: tracks
    with 10 OMs}
\label{fig:pul2}
\end{figure}

The quality of the reconstruction can be assessed by the usual
criteria of the pull distribution and the $\chi^2$ probability
distribution. The latter is presented in fig. \ref{fig:chiprob}, while
the pulls for $\theta$ and $\phi$ in fig. \ref{fig:pul}.
These plots were produced with the reconstruction using the  {\em Constant deweight}
method (see \S \ref{sec:fit}), with a deweight parameter equal to 6
(see below). For completeness, we include here the
$\sigma$ values for the gaussian fits of these
distributions:
\begin{itemize}
\item $\phi$ distribution (fig. \ref{fig:phipul}):
  \begin{itemize}
  \item When the $\phi$ distribution without
    cuts (fig. \ref{fig:phipul}, left) is fitted,  $\sigma =
    1.549$. If only the peak of the distribution is fitted (from $-2$
    to $2$), then $\sigma =1.097$.
  \item When the $\phi$ distribution with a cut at 10 OMs
    (fig. \ref{fig:phipul}, right) is fitted,  $\sigma =
    1.162$. If only the peak of the distribution is fitted (from $-2$
    to $2$), then $\sigma =1.024$.
  \end{itemize}
\item $\theta$ distribution (fig. \ref{fig:thetapul}):
  \begin{itemize}
  \item   When the $\theta$ distribution without
    cuts (fig. \ref{fig:thetapul}, left) is fitted,$\sigma  =
    1.615$. $\sigma$.  If only the peak of the distribution is fitted (from $-2$
    to $2$), then $\sigma = 1.161$.
  \item When the $\theta$ distribution with a cut at 10 OMs
    (fig. \ref{fig:thetapul}, right) is fitted,  $\sigma =
    1.209$. If only the peak of the distribution is fitted (from $-2$
    to $2$), then $\sigma =1.066$.
  \end{itemize}
\end{itemize}
The 
naturally ascociated error  with each
hit ($\sigma_i$ in eq. \ref{eq:chi2}) is initially set equal with the time
uncertainty of 2 ns (more or less equal with the time resolution
available). For these plots the hits were reweighted with the rather
large value of 6 ns, and the justification can be seen in plots
\ref{fig:chiprob2} and \ref{fig:pul2}, which were produced with an
error value of 2 ns. The sharp peak at 0 of the $\chi^2$ plot
\ref{fig:chiprob2}, shows that we underestimate the error.
This fact is 
also reflected to the 
pull distributions (figs. \ref{fig:pul2})  whose $\sigma$ is about
3 or 4, as opposed to figs. \ref{fig:pul}. This in turn, is a reflection of the
highly non-gaussian nature of the errors ascociated with the problem,
due mainly  to the multitude of different interactions involved.
For example,  the  number of photons emitted from a track and the form of their
distribution in space and time, when brehmsstralung and hadronic proceses are involved, is
drastically different than the one expected from the simple formula
used, eq. \ref{eq:geom}, which assumes that all hits are prompt photons.

Therefore, the choice was made to use a larger than expected error for
the fitting process, 
since then the pulls' $\sigma$  becomes 1. 
The
application of a reasonable cut (such as 10 OMs), diminishes the tails also,
which means that in this case the errors on the track parameters as
calulated by the fit are more representative of the
true situation (see fig.  \ref{fig:pul}). This, in short, is the reason
that a large choice for the fitting error is more adequate. 

\begin{figure}[htpb]
  \centering
  \includegraphics[width=120mm]{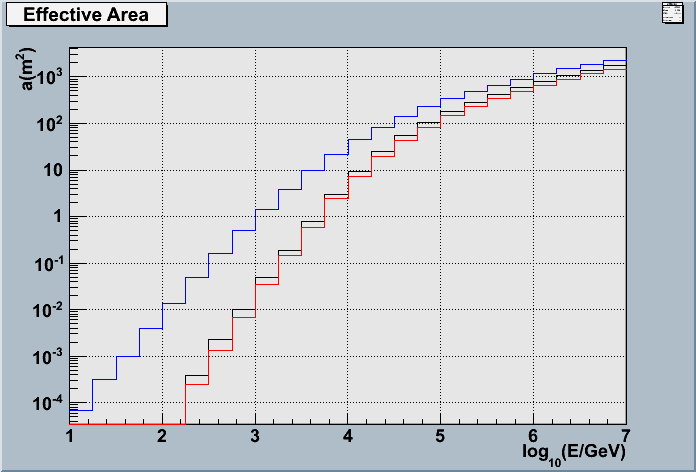}
  \caption{$\nu$ effective areas. Blue: all reconstructed events. Black: only events with
    at least 10 OMs. Red: events with
    at least 10 OMs, reconstructed within $5^{\circ}$ from the MC direction. }
\label{fig:nueff}
\end{figure}

The neutrino effective area vs energy is plotted in
fig. \ref{fig:nueff}, where we compare the effective area for all
reconstructed tracks (blue line), to the effective area when only
those tracks with at least 10 OMs are taken into account (black line). 
As a measure of the real efficiency of the reconstruction, we also
plot (the red line)  the effective area {\em without} the fakes (i.e. those tracks
that are reconstructed more that $5^{\circ}$ away from the true direction).

\begin{figure}[htpb]
  \centering
  \includegraphics[width=120mm]{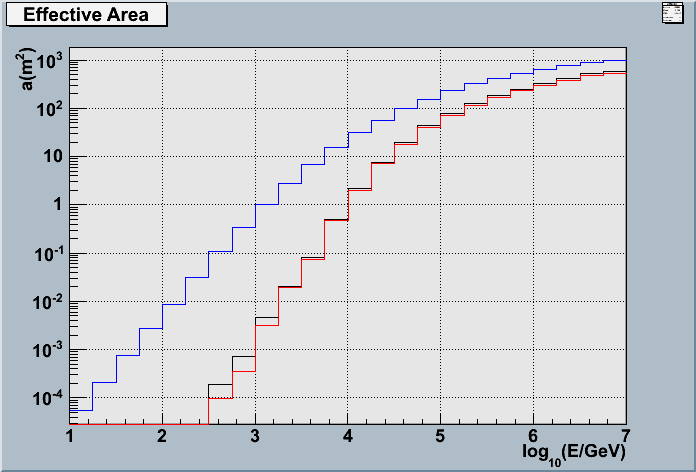}
  \caption{Same as \ref{fig:nueff}, but only for upgoing muons
    ($z>90^{\circ}$).}
\label{fig:nueffup}
\end{figure}



\chapter{Depth Studies}

The measure of the quality of a detector is its
``sensitivity'' \cite{hill}, and here we expore its relation to
depth. The procedure for the calculation of the
sensitivity is based on the Feldman-Cousins technique \cite{feld}. 

The estimation of the atmospheric $\nu$ backgrounds was based
on the MC files, described already in \S \ref{sec:data}. The
atmospheric $\mu$ backgrounds on the other hand, pose a difficult
problem, due mainly to their large numbers -- and the resulting need
for resources.  Within the present work, a rough MC estimation was
done for them, while the more quantitative full Monte Carlo study will
follow soon. 

After a short description of the method we use for
calculating the fluxes in the first section of this chapter, 
the results are presented in \S \ref{sec:sens}.

\section{Fluxes}

The first step in calculating the point source sensitivity, is the
evaluation of atmospheric and astrophysical  $\nu$ fluxes.  
The outline of the calculation is as follows: 
\begin{itemize}
\item Extract the events' zenith and azimuth. Calculate declinations
  and right ascensions (see \S \ref{sec:sens} for a discussion of
  some relevant technical details and fig. \ref{fig:aitoff} for a sky map). 

\item Extract the weights, as calculated from the simulation (for an
  explanation see \cite{weight}).
\item Calculate the signal rate $R_i ^s$ as 
  \begin{equation}
    \label{eq:sigrate}
    R_i ^s = \Phi_{\imath}^s \frac{O_{\imath}}{N} T,
  \end{equation}
where $N$ the total number of events, $O_{\imath}$ the {\tt OneWeight}
parameter, $T$ the live time in seconds 
and
\begin{equation}
  \label{eq:lfux}
  \Phi_i^s = \frac{2.6\times 10^{-8}}{E_{\imath}^2}
\end{equation}
is the signal flux.
\begin{figure}[htb]
  \centering
\includegraphics[width=70mm]{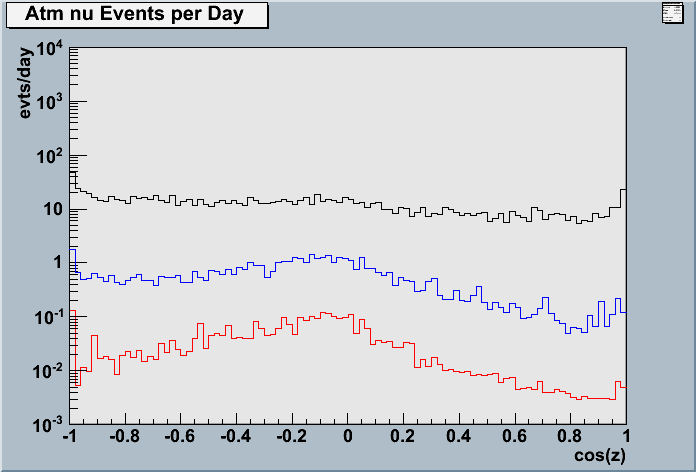}
  \caption{Expected Atm. neutrino rate (as per
eq. \ref{eq:bgrate}). The nature of the peaks at $\pm 1$ can be, at least
partially, attributed to ghosts, see \S \ref{sub:asymmetry}. Black: All
reconstructed    events. Blue: Reconstructed with at least 6 OMs. Red:
Reconstructed with at least 10 OMs.}
  \label{fig:rates}
\end{figure} 
\item Calculate the relevant background rate as 
  \begin{equation}
    \label{eq:bgrate}
   R_i ^b = \Phi_{\imath}^b \frac{O_{\imath}}{N} 2T,
  \end{equation}
where now $\Phi_{\imath}^b$ is a bartol flux \cite{geisser} (see \S \ref{sec:data} and fig. \ref{fig:rates}). 
The factor of $2$ is due to the counting of both $\nu$s and
$\bar\nu$s. 

\end{itemize}

All the above factors are related to the irreducible background of
atmospheric neutrinos, which is practically unchanged with depth. The
only factor that depends on depth is the flux of atmospheric
muons. Due to limitations of time, the full Monte Carlo study of
atmospheric muon fluxes will be performed at a later stage.

From the user point of view, the task is divided in two parts. 
The first part is implemented  in
the {\tt EffAreaPlots} module, which
outputs the data needed for the calculation to a chain of ROOT files. The
second part of the calculation is finished by a series of scripts
manipulating the data
(see esp. {\tt EffAreaPlotsMaker} and the functions therein, and the
relevant implementation scripts, e.g. {\tt analysis\_plots.C} and {\tt effAreapyROOT.py})

\section{Sensitivity}
\label{sec:sens}

We calculate the point source sensitivity by using the 
straightforward  fixed bin method, where we use an angular bin of the
order of the angular resolution  
(For a basic definition of terms see \cite{ebetza}).
The main idea is that signal events will cluster around the direction
of the astrophysical neutrino source under consideration (within
of course the angular resolution of the detector), therefore 
producing an excess of events over the uniform background. 

The background is made of two parts: the atmospheric neutrino part
and the atmospheric muon part. The former is already included from the
MC source files that were reconstructed. 
For the latter, a full reconstruction is not easy to be performed
 because of time and CPU
constraints (work is underway in this direction). 
Until the inclusion of the full  Monte Carlo data for the 
atmospheric muon
fluxes, a general argument can be used: 
On one hand, the atmospheric muon intensity, as a function of sea depth, is a well
documented quantity (see e.g. \cite{cdr}, (fig. 1-3) and discussion
thereafter, or \cite{muonint}, fig. 15). On the other, the number of expected atmospheric muons is
already given from the existing simulations.
From a preliminary study of atmospheric muons at 3500 m, using  the
program {\tt mupage}, the estimated muons that reach the detector
would be of the order of $8\times 10^9$ events per year.
This number is corroborated by (fig. 1-3) of  \cite{cdr}, for a
detector of approximate surface $1 km^2$.

The detector produces fakes (here the word ``fakes'' signifies
downgoing events that were
misreconstructed as upcoming) at a known percentage of input events.
In fig. \ref{fig:fakepercent} the ratio of misreconstructed events
over the total produced downgoing MC neutrino events is shown. This ratio can be used for
an evaluation of the number of misreconstructed atmospheric muons,
since their energy spectrum, as a function of depth, is known. The
result of this exercise is shown in fig. \ref{fig:misreco}.  
\begin{figure}[htp]
  \centering
\label{fig:atmspect}
\includegraphics[width=70mm]{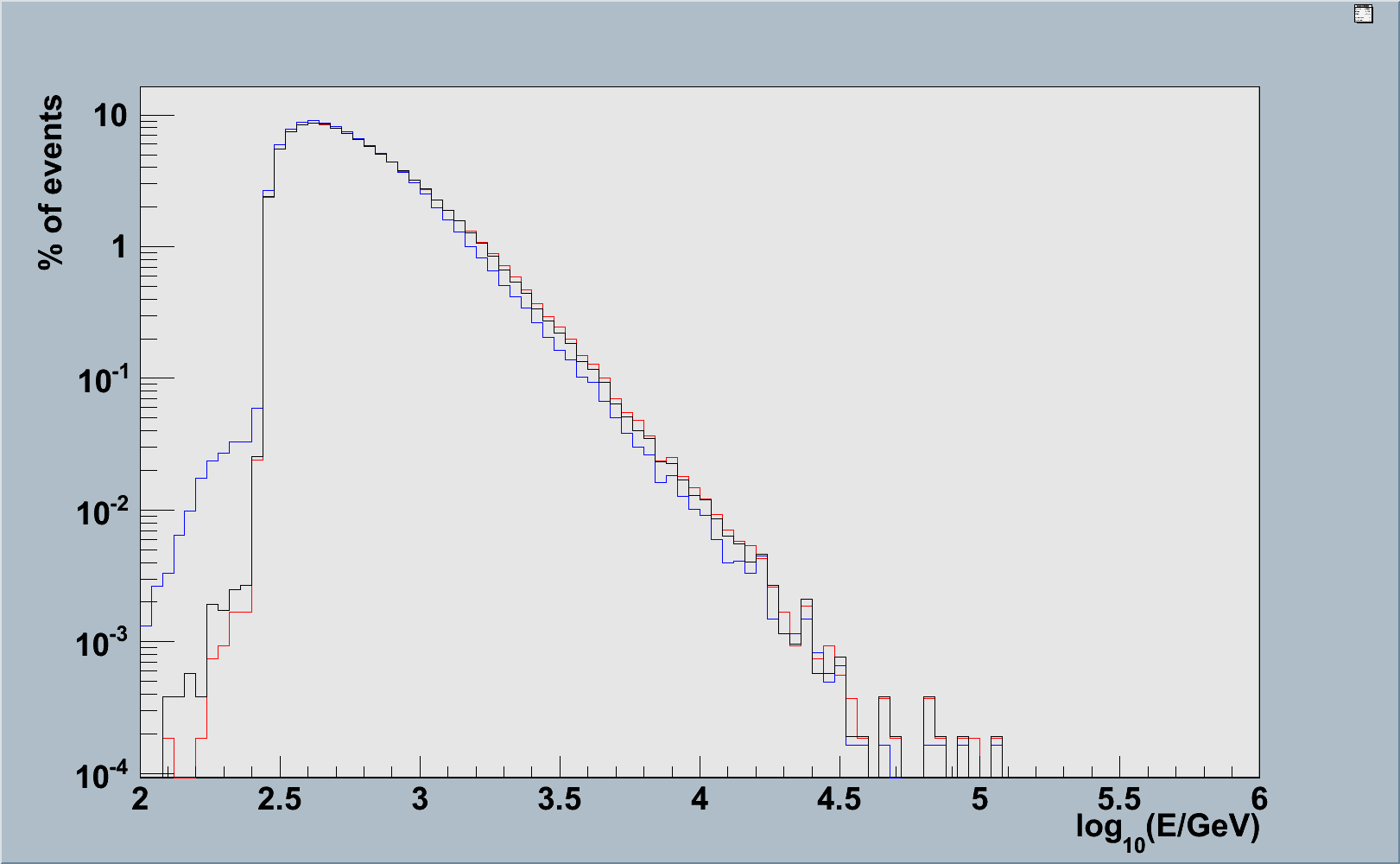}
\caption{Energy spectrum of atm $\mu$'s. Red: Depth = 5000m. Black: Depth =
  4500 m. Blue: Depth=3500 m.}
\end{figure} 
\begin{figure}[htp]
  \centering
  \subfloat[Ratio of misreconstructed ``fakes'' over total number of MC
  events.]
  {\label{fig:fakepercent}
    \includegraphics[width=70mm]{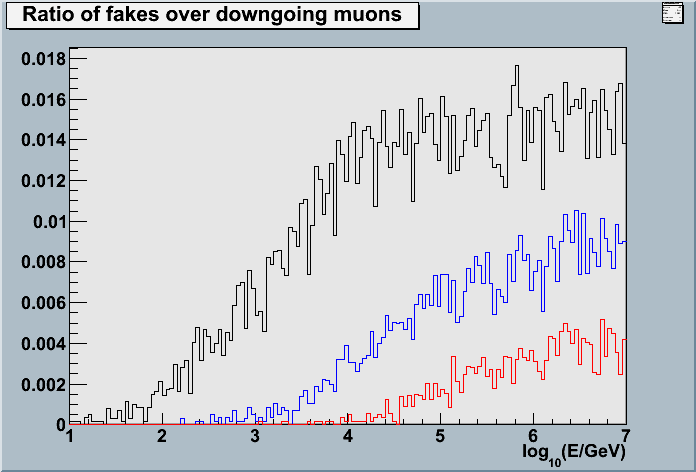}}
  \quad
  \subfloat[Expected atm muon ``fakes'' over one year   period.    ]
  {\label{fig:misreco} 
    \includegraphics[width=70mm]{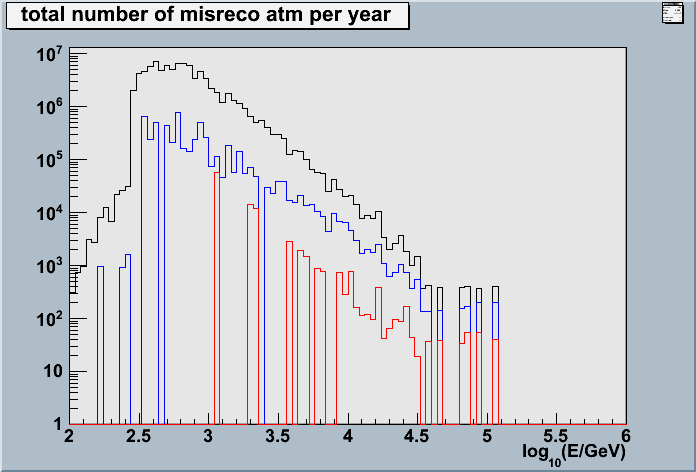}}
\caption{Black: All reconstructed events. Blue: Reconstructed with
  at least 6 OMs. Red: Reconstructed with at least 10 OMs.}
\end{figure}

The total number of estimated fakes, at 10 OMs cut, is 94754, for a
live time of one year. For a uniform distribution of the fakes over the
whole lower part of the sky and  an angular bin of radius $0.4^{\circ}$, this
is reduced to 2.3 fakes per bin per year.
It should be noted that, due to low statistics, the number of fakes is overestimated
slightly (observe the fluctuations around $10^3 GeV$ in
fig. \ref{fig:misreco}), and averaged over the whole sky, something that
is expected to be amended in the near future, after a complete
reconstruction of the full data set.

As a comparison to the above number, the mean number of expected atmospheric $\nu$ background
events per angular bin of $0.4^{\circ}$ per year is 4.4.

These numbers are in good general agreement with the relevant  MC simulations published so
far: see e.g. fig. \ref{fig:ernplot}, where the atmospheric muon intensity at 3500 m is of the same order of
magnitude as the neutrino background.
As a qualitative rule then, one can extrapolate the information about
the atmospheric background from its two sources, calculating the
total background (Atm $\mu$'s + Atm $\nu$'s), as follows:
If at 3500 m, the $\mu$-background is expected to be on the average
half the $\nu$ background, then at 2500 m, their ratio is expected to
be roughly 2, and conversely at 4500 it will be 0.11 and at 5200 0.04.

The plot for the sensitivity for a cone of $0.4^{\circ}$ around the MC
direction is shown in fig. \ref{fig:sensitive}. A calculation of the
sensitivity of our detector for various $\mu$-background scenarios and
for 2 different cuts on the number of OMs is presented
in Table \ref{arr:sens}, for a declination of $53^{\circ}$, near the
minimum. 
The left column is the ratio of the
atm-$\mu$ background over the $\nu$ background, i.e. when
this quantity is 1, there are equal numbers of $\nu$ and $\mu$
background events. 

\begin{table}[h!t!p!]
\centering

\caption{Sensitivities $(GeV^{-1} cm^{-2} s^{-1}  \times 10^{-10})$, with an estimated atm $\mu$ background. }
\begin{tabular}{|c|c|c|}

\hline
&&\\
atm $\mu$ backg  & sensitivity, $90 \%$ CL& sensitivity, $90 \%$ CL\\
above atm $\nu$& {\em 6 OMs} cut &{\em 10 OMs} cut\\[5pt]
\hline
&&\\[-9pt]
0     &  6.06449  &  7.84659     \\
 0.5  &  6.38169  &  7.95273  \\
 1.0  &  6.74396  &  8.05637  \\
 1.5  &  7.0803   &  8.16001  \\
 2.0  &  7.38752  &  8.26084  \\
 2.5  &  7.67674  &  8.36077  \\
 3.0  &  7.94508  &  8.46023  \\
 3.5  &  8.20192  &  8.55668  \\
 4.0  &  8.43783  &  8.65312  \\
 5.0  &  8.87264  &  8.84108  \\
 6.0  &  9.27673  &  9.02472  \\
 7.0  &  9.67098  &  9.21145  \\
 8.0  &  10.0188  &  9.43918  \\
 9.0  &  10.3432  &  9.64955  \\
 20.0  &  13.4414  &  11.5359  \\
 50.0  &  19.335   &  14.9283  \\
100.0  &  25.9965  &  18.9767  \\

\hline

\end{tabular}
\label{arr:sens}
\end{table}

 \begin{figure}[htp]
  \centering
  \includegraphics[width=140mm]{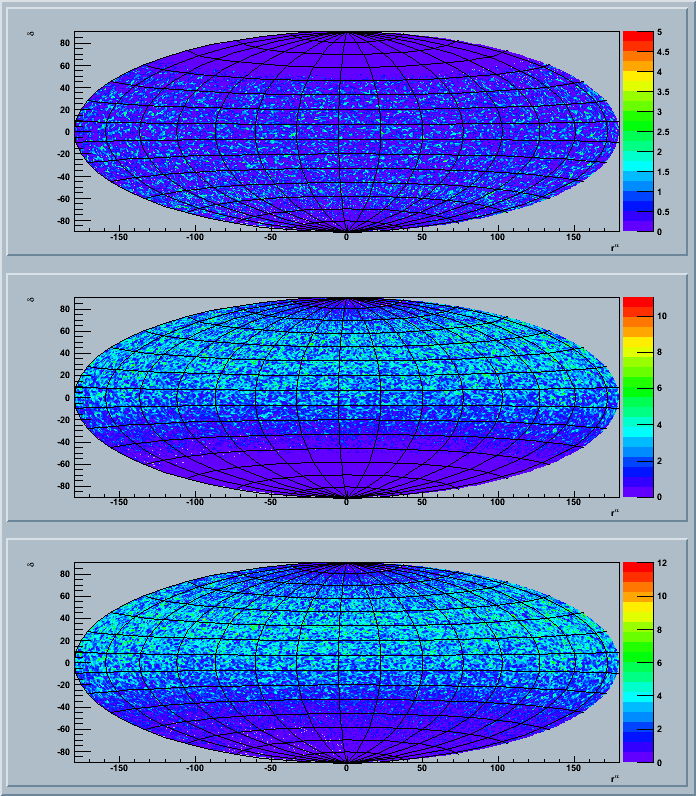}
  \caption{A sky map of reconstructed  muons in equitorial
    coordinates declination $\delta$
    and right ascension $r^{\alpha}$.
    Upper: locally upcoming muons ($z>90^{\circ}$).
    Middle: downgoing muons ($0^{\circ} \leq z <90^{\circ}$).
    Lower: Total. Not a very useful plot, but it does display nice colors.}
  \label{fig:aitoff}
\end{figure}


\begin{figure}[htp]
  \centering
  \includegraphics[width=160mm]{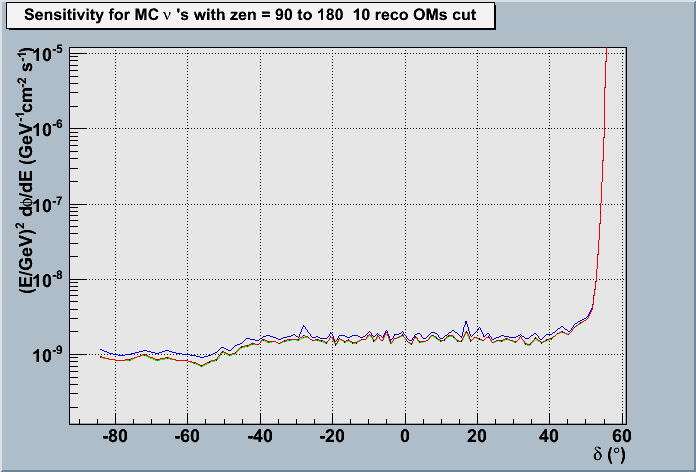}
  \caption{One year sensitivity at the Nestor site, for a cut at 10 OMs.  Green line: no atm
    background. Red Line: An estimate for atm $mu$ background equal to
    atm $\nu$ background. Blue line: An atm $mu$ background equal to
    10 times the atm $\nu$ background. For a discussion see section \ref{sec:sens}.}
  \label{fig:sensitive}
\end{figure}

\pagebreak
\newpage

\begin{figure}[t!]
  \centering
  \includegraphics[width=60mm]{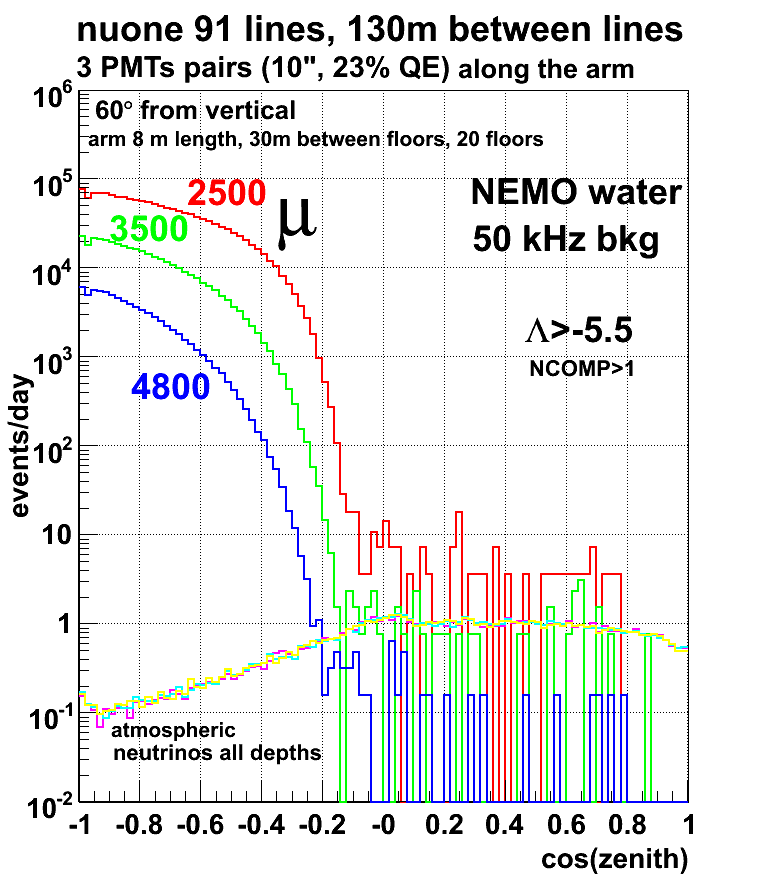}
  \caption{ Taken from \cite{ernPlot}.}
  \label{fig:ernplot}
\end{figure}




\end{document}